\documentclass[prd,superscriptaddress,twocolumn,preprintnumbers,showpacs,nofootinbib]{revtex4-1} 
\pdfoutput=1
\usepackage{amssymb,amsmath,graphicx,epsfig}
\usepackage{hyperref,tabu,color,array}
\usepackage{float,url}
\usepackage{hyperref}
\usepackage{slashed}
\usepackage[titletoc,title]{appendix} 
\usepackage[export]{adjustbox}
\hypersetup{linktocpage=true}
\hypersetup{
colorlinks=true,   % false:	 links; true:colored links
linkcolor=blue, % color of internal links
citecolor=blue, % color of links to bibliography
filecolor=blue,    % color of file links
urlcolor=blue      % color of external links
}
\def\beq{\begin{equation}} 
\def\eeq{\end{equation}} 
\def\bea{\begin{eqnarray}} 
\def\eea{\end{eqnarray}} 
\def\nn{\nonumber}

\def\cNINEe{[\Delta C_{9}^e]}
\def\cNINEpe{[\Delta C_{9'}^e]}
\def\cOTENe{[\Delta C_{10}^e]}
\def\cOTENpe{[\Delta C_{10'}^e]} 
\def\cSe{[\Delta C_{S}^e]}
\def\cSpe{[\Delta C_{S'}^e]}
\def\cssPe{[\Delta C_{P}^e]}
\def\cssPpe{[\Delta C_{P'}^e]}
\def\cTe{[\Delta C_{T}^e]}
\def\cTFIVEe{[\Delta C_{T5}^e]}
\def\cNINEmu{[\Delta C_{9}^\mu]}
\def\cNINEpmu{[\Delta C_{9'}^\mu]}
\def\cOTENmu{[\Delta C_{10}^\mu]}
\def\cOTENpmu{[\Delta C_{10'}^\mu]} 
\def\cSmu{[\Delta C_{S}^\mu]}
\def\cSpmu{[\Delta C_{S'}^\mu]}
\def\cssPmu{[\Delta C_{P}^\mu]}
\def\cssPpmu{[\Delta C_{P'}^\mu]}
\def\cTmu{[\Delta C_{T}^\mu]}
\def\cTFIVEmu{[\Delta C_{T5}^\mu]}
\def\cNINEell{[\Delta C_{9}^\ell]}
\def\cNINEpell{[\Delta C_{9'}^\ell]}
\def\cOTENell{[\Delta C_{10}^\ell]}
\def\cOTENpell{[\Delta C_{10'}^\ell]} 
\def\cSell{[\Delta C_{S}^\ell]}
\def\cSpell{[\Delta C_{S'}^\ell]}
\def\cssPell{[\Delta C_{P}^\ell]}
\def\cssPpell{[\Delta C_{P'}^\ell]}
\def\cTell{[\Delta C_{T}^\ell]}
\def\cTFIVEell{[\Delta C_{T5}^\ell]}

\begin{document} 
\title{Role of Tensor operators in $R_K$ and $R_{K^*}$}
%%%%%%%%%%%%%%%%%%%%%%%%%%%%%%%%%%%%%%%%%%%%%%%%%%%%%%%%%%%%%%%%%%%%%%%%%%%%%%% 
\author{Debjyoti Bardhan} 
\email{debjyotiarr@gmail.com} 
\affiliation{\normalfont{Department of Theoretical Physics, Tata Institute of Fundamental Research, 1 Homi Bhabha Road, Mumbai 400005, India}}
\author{Pritibhajan Byakti}
\email{tppb@iacs.res.in}
\affiliation{\normalfont{Department of Theoretical Physics, Indian Association for the Cultivation of Science, 2A \& 2B, Raja S.C. Mullick Road, 
Jadavpur, Kolkata 700 032, India}}
\author{Diptimoy Ghosh}
\email{diptimoy.ghosh@weizmann.ac.il} 
\affiliation{\normalfont{Department of Particle Physics and Astrophysics, Weizmann Institute of Science, Rehovot 76100, Israel}}
%%%%%%%%%%%%%%%%%%%%%%%%%%%%%%%%%%%%%%%%%%%%%%%%%%%%%%%%%%%%%%%%%%%%%%%%%%%%%%% 
\begin{abstract} 
The recent LHCb measurement of $R_{K^*}$ in two $q^2$ bins, when combined with the earlier measurement of $R_K$, strongly suggests 
lepton flavour non-universal new physics in semi-leptonic $B$ meson decays. Motivated by these intriguing hints of new physics, several 
authors have considered vector, axial vector, scalar and pseudo scalar operators as possible explanations of these measurements. However, 
tensor operators have widely been neglected in this context. In this paper, we consider the effect of tensor operators in $R_K$ and $R_{K^*}$. 
We find that, unlike other local operators, tensor operators can comfortably produce both of $R_{K^*} ^{\rm low}$ and $R_{K^*} ^{\rm central}$ 
close to their experimental central values. However, a simultaneous explanation of $R_K$ is not possible with only Tensor operators, and other 
vector or axial vector operators are needed. In fact, we find that combination of vector and tensor operators can provide simultaneous explanations 
of all the anomalies comfortably at the $1 \sigma$ level, a scenario which is hard to achieve with only vector or axial vector operators. 
We also comment on the compatibility of the various new physics solutions with the measurements of the inclusive decay $B_d \to X_s \ell^+ \ell^-$. 
\end{abstract}
%%%%%%%%%%%%%%%%%%%%%%%%%%%%%%%%%%%%%%%%%%%%%%%%%%%%%%%%%%%%%%%%%%%%%%%%%%%%%%%
%\keywords{}
%\pacs{14.40.Nd, 13.20.He, 12.60.Cn}
%\preprint{}
\maketitle
%14.40.Nd   Bottom mesons (|B| > 0) 
%13.20.He   Decays of bottom mesons 
%12.60.Cn   Extensions of electroweak gauge sector
%%%%%%%%%%%%%%%%%%%%%%%%%%%%%%%%%%%%%%%%%%%%%%%%%%%%%%%%%%%%%%%%%%%%%%%%%%%%%%%
\subsection{\rm \bf Introduction}
%\vspace*{-3mm}
%%%%%%%%%%%%%%%%%%%%%%%%%%%%%%%%%%%%%%%%%%%%%%%%%%%%%%%%%%%%%%%%%%%%%%%%%%%%%%%
The LHCb collaboration has recently announced measurements of 
$R_{K^*} \equiv  {\cal B}({\bar B}_d \to \bar K^* \mu^+ \mu^-)/{\cal B}({\bar B}_d \to \bar K^* e^+ e^-)$ in two $q^2 (\equiv (p_{\ell^+} + p_{\ell^-})^2)$ bins, 
$[0.045,\, 1.1]$ and $[1.1,\, 6]$ GeV$^2$ (referred to as low and central bins respectively) \cite{Aaij:2017vbb}. In both the bins, they observe deviation from the Standard Model (SM), at 
the $2.1-2.3 \sigma$  level in the low bin and at the $2.4-2.5 \sigma$ level in the central bin \cite{Aaij:2017vbb}. Interestingly, in the summer of 
2014, a similar LHCb measurement of the ratio 
$R_{K} \equiv  {\cal B}(B^+ \to K^+ \mu^+ \mu^-)/{\cal B}(B^+ \to K^+ e^+ e^-)$ for $q^2 \in [1,6]$ GeV$^2$ also showed a $2.6 \sigma$ 
deviation from the SM \cite{Aaij:2014ora}.  The experimental measurements as well as the latest SM predictions for these ratios are summarised in 
the first 3 rows of Table-\ref{tab:exp-data}.

As the theoretical predictions of $R_{K}$ and $R_{K^*}$ in the SM are rather reliable \cite{Hiller:2003js,Bordone:2016gaq}, these measurements highly suggest 
for lepton non-universal new physics (NP). This has spurred a lot of activities in the recent past, both in the language of model independent higher 
dimensional operators and specific models beyond the SM \cite{Hiller:2003js,Altmannshofer:2008dz,Alok:2009tz,Alok:2010zd,Alok:2011gv,
DescotesGenon:2011yn,Altmannshofer:2011gn,Matias:2012xw,DescotesGenon:2012zf,Lyon:2013gba,Descotes-Genon:2013wba,
Altmannshofer:2013foa,Buras:2013qja,Datta:2013kja,Altmannshofer:2014cfa,Alonso:2014csa,Ghosh:2014awa,Queiroz:2014pra,Mandal:2014kma,Gripaios:2014tna,Greljo:2015mma,
Gripaios:2015gra,Alonso:2015sja,Barbieri:2015yvd,Falkowski:2015zwa,Crivellin:2015lwa,Crivellin:2015era,Belanger:2015nma,Carmona:2015ena,Mandal:2015bsa,Sahoo:2015wya,Sahoo:2016pet,Crivellin:2016ejn,Feruglio:2016gvd,Barbieri:2016las,GarciaGarcia:2016nvr,Megias:2016bde,
Bhattacharya:2016mcc,Bhatia:2017tgo,Megias:2017ove,Datta:2017pfz,Bordone:2017anc,DiChiara:2017cjq,
%%%%%%%%%
Capdevila:2017bsm,Altmannshofer:2017yso,DAmico:2017mtc,Hiller:2017bzc,
Geng:2017svp,Ciuchini:2017mik,Celis:2017doq,Becirevic:2017jtw,Ghosh:2017ber,Alok:2017jaf,Alok:2017sui,Wang:2017mrd,Feruglio:2017rjo,
Ellis:2017nrp,Alonso:2017bff,Bishara:2017pje,Alonso:2017uky,Tang:2017gkz,Datta:2017ezo,Das:2017kfo}. 
In the context of dimension-6 NP operators, it has been pointed out that short distance NP operators 
of certain types can provide an overall good fit to the data. However, a discussion of the tensor operators was missing. 
In this paper, we fill this gap with a detailed analysis of the role of tensor operators in $R_K$ and $R_{K^*}$\footnote{In the context of 
${\bar B}_d \to {\bar K}^* \ell^+ \ell^-$ decay, the tensor operators with $m_\ell \neq 0$ was first considered by one of the authors in 
\cite{Alok:2009tz,Alok:2010zd,Alok:2011gv} and later in \cite{Bobeth:2012vn,Gratrex:2015hna}.}.

Note that, it is not possible to generate tensor operators at the dimension-6 level if the Standard Model gauge symmetry is 
imposed \cite{Alonso:2014csa}. However, tensor operators can be generated at the dimension-8 level, see the end of 
section~\ref{combination} for more details.
%
%%%%%%%%%%%%%%%%%%%%%%%%%%%%%%%%%%%%%%%%%%%%%%%%%%%%%%%%%%%%%%%%%%%%%%%%%%%%%%%dimension-8
\begin{table}[ht!]
\begin{center}
\begin{tabular}{|l|cr|cr|} 
\hline
Observable & SM prediction &  & Measurement  &  \\[1mm]
\hline
%%%%%
$R_K^{\rm cen}$ & $1.00 \pm 0.01 $&  \cite{Descotes-Genon:2015uva,Bordone:2016gaq} &  $[0.66, \, 0.84]$  & \cite{Aaij:2014ora} \\[1mm]
\hline
%%%%%
$R_{K^*} ^{\rm low}$  & $0.92 \pm 0.02$  &  \cite{Capdevila:2017bsm} &  $[0.58,\, 0.77]$  & \cite{Aaij:2017vbb} \\[1mm]
\hline
%%%%%
$R_{K^*}^{\rm cen}$  & $1.00 \pm 0.01 $&  \cite{Descotes-Genon:2015uva,Bordone:2016gaq} & $[0.60,\,0.81]$  & \cite{Aaij:2017vbb} \\[1mm]
\hline
%%%%%
$\mathcal{B}_{\mu\mu} \times 10^{9}$ & $3.57 \pm 0.16$ & \cite{Bobeth:2013uxa,Fleischer:2017ltw}
& $[2.5,\,3.5]$ & \cite{Chatrchyan:2013bka,Aaij:2017vad,Fleischer:2017ltw} \\[1mm]
\hline
%%%%%
$\mathcal{B}_{ee} \times 10^{14}$ & $8.35 \pm 0.39$ & \cite{Bobeth:2013uxa,Fleischer:2017ltw}
& $< 2.8 \times 10^7$ & \cite{Aaltonen:2009vr} \\[1mm]
\hline
%%%%%
$\mathcal{B}^{\text{low}}_{X_s\mu\mu} \times 10^{6}$ & $1.59 \pm 0.11$   & \cite{Huber:2007vv}  & $[0,\,1.53]$   & \cite{Lees:2013nxa} \\[1mm]
\hline
%%%%%
$\mathcal{B}^{\text{high}}_{X_s\mu\mu} \times 10^{6}$ & $0.24 \pm 0.07$   &  \cite{Huber:2007vv}  & $[0.31,\,0.91]$  &  \cite{Lees:2013nxa} \\[1mm]
\hline
%%%%%
%%%%%
$\mathcal{B}^{\text{low}}_{X_see} \times 10^{6}$ & $1.64 \pm 0.11$   & \cite{Huber:2007vv}   &  $[1.42,\,2.47]$ & \cite{Lees:2013nxa}  \\[1mm]
\hline
%%%%%
$\mathcal{B}^{\text{high}}_{X_see} \times 10^{6}$ & $0.21 \pm 0.07$  & \cite{Huber:2007vv}   & $[0.38,\,0.75]$  & \cite{Lees:2013nxa} \\[1mm]
\hline
%%%%%
\hline
\end{tabular}
\caption{Observables, their SM predictions and experimental $1\sigma$ ranges. For $R_{K^*} ^{\rm low}$, a more conservative SM prediction,  
$0.906 \pm 0.028$, has been recently reported in \cite{Bordone:2016gaq}.
\label{tab:exp-data}}
\end{center}
\end{table} 
%%%%%%%%%%%%%%%%%%%%%%%%%%%%%%%%%%%%%%%%%%%%%%%%%%%%%%%%%%%%%%%%%%%%%%%%%%%%%%%

Besides $R_K$ and $R_{K^*}$, we also consider the branching ratios of $B_s \to \ell^+ \ell^- (\ell = \mu, e)$ as they are reliably predicted in the 
SM. Furthermore, we also show the compatibility with measurements of the branching ratios of the inclusive decay $B_d \to X_s \ell^+ \ell^-$. 
The experimental measurements of these observables are summarised in Table-\ref{tab:exp-data}. In the table and the subsequent text, we use the 
following short-hand notations ($q^2$ is given in GeV$^2$)
\bea
&&{\cal B}_{K\ell\ell}^{\rm cen} \equiv {\cal B}(B^+ \to K^+ \ell^+ \ell^-), \,  q^2 \in [1,\, 6]  \nn \\
&&{\cal B}_{K^*\ell\ell}^{\rm low(cen)} \equiv {\cal B}({\bar B}_d \to {\bar K}^* \ell^+ \ell^-), \,  q^2 \in [0.045,\, 1.1]\, ( [1.1,\, 6] ) \nn \\
&&\mathcal{B}_{\ell\ell} \equiv {\cal B}(\bar B_s \to \ell^+ \ell^-) \nn \\
&&{\cal B}_{X_s\ell\ell}^{\rm low (high)} \equiv {\cal B}({\bar B}_d \to X_s \ell^+ \ell^-), \,  q^2 \in [1,\, 6] \, ( [14.2,\, 25] ) \nn
\eea
We will not consider any angular observables ($P_5^\prime$, for example) in this analysis because their SM predictions are debatable 
\cite{Ball:2006eu,Khodjamirian:2010vf,Dimou:2012un,Jager:2012uw,Lyon:2013gba,Lyon:2014hpa,Ciuchini:2015qxb,Jager:2014rwa,
Capdevila:2017ert}
\footnote{Note however that, large deviations from the SM expectations in two $q^2$-bins of $P_5^{\prime\, \mu}$ have been 
claimed in the literature \cite{Descotes-Genon:2013wba}. Interestingly, the Belle collaboration has provided the first measurement 
of $P_5^{\prime}$ in the electron mode \cite{Wehle:2016yoi}, and indeed, the central value for $P_5^{\prime\, \mu}$ deviates 
more than that of $P_5^{\prime\, e}$. However, at this point the statistics is low, and the jury is still out on this.}.
In our calculations of ${\cal B}({\bar B}_d \to {\bar K}^{(*)} \ell^+ \ell^-)$ we only include the factorizable part described by the form-factors, 
and no non-factorizable corrections are included. However, this is good enough for the theoretically clean observables $R_K$ and $R_{K^*}$. 
As for the form-factors, we use \cite{Bouchard:2013pna} for $B \to K$ matrix elements and \cite{Straub:2015ica} for the $B \to K^*$ matrix elements.

%%%%%%%%%%%%%%%%%%%%%%%%%%%%%%%%%%%%%%%%%%%%%%%%%%%%%%%%%%%%%%%%%%%%%%%%%%%%%%%
%\vspace*{-5mm}
\subsection{\rm \bf Effective~operators}
\label{sec:eff_ops}
%\vspace*{-2mm}
%%%%%%%%%%%%%%%%%%%%%%%%%%%%%%%%%%%%%%%%%%%%%%%%%%%%%%%%%%%%%%%%%%%%%%%%%%%%%%%

The $\rm SU(3) \times U(1)$ invariant effective Lagrangian at the dimension-6 level for $b \to s$ transition is given by 
\begin{align} 
\label{eq:heff}
{\cal L}_{\mathrm{eff}} = -\frac{4G_F}{\sqrt{2}} \, \frac{\alpha_{\rm em}}{4 \pi} \, V_{tb} V_{ts}^{*}  \, {\cal{H}}_{\mathrm{eff}}^{(t)}  + \rm h.c.
\end{align}
\vspace{-5mm}
\begin{eqnarray}
\text{where, }{\cal{H}}_{\mathrm{eff}}^{(t)} = C_1 {\cal{O}}_1^c + C_2 {\cal{O}}_2^c + \sum_{i=3}^{6} C_i {\cal{O}}_i + 
\sum_{i=7}^{10} C_i {\cal{O}}_i \nn
 \end{eqnarray}
In models beyond the SM, new operators can be generated. The complete basis of dimension-6 operators 
includes new operators given by 
\begin{align}
{\cal{H}}_{\mathrm{eff}}^{(t), \, \rm New} &=  \sum_{i=7, 9, 10} C_{i^\prime} {\cal{O}}_{i^\prime} + 
\hspace{-2mm} \sum_{i=S, P, S', P', T, T5}  \hspace{-5mm} C_i {\cal{O}}_i
\end{align}
where, the various operators above are defined by
\begin{align}
\mathcal{O}_{7(7')}&=\frac{1}{e} \, m_b [\overline{s}\sigma_{\mu\nu} P_{R(L)} b] F^{\mu\nu} \nonumber \\ 
\mathcal{O}_{9(9')}&=[\overline{s}\gamma_\mu P_{L(R)} b][\overline{l}\gamma^\mu l], \, 
\mathcal{O}_{10(10')}=[\overline{s}\gamma_\mu P_{L(R)} b][\overline{l}\gamma^\mu \gamma_5 l] \nonumber \\ 
\mathcal{O}_{S(S')}&=[\overline{s} P_{R(L)} b] \, [\overline{l} l], \, 
\hspace{6.8mm} \mathcal{O}_{P(P')}=[\overline{s} P_{R(L)} b][\overline{l} \gamma_5 l] \nonumber \\
\mathcal{O}_T&=[\overline{s} \sigma_{\mu\nu} b][\overline{l} \sigma^{\mu\nu} l], \, 
\hspace{10.4mm} \mathcal{O}_{T5}=[\overline{s} \sigma_{\mu\nu} b][\overline{l} \sigma^{\mu\nu} \gamma_5 l] \nonumber 
\end{align}

Note that, the Wilson coefficients for the photonic dipole operators $\mathcal{O}_{7}$ and $\mathcal{O}_{7^\prime}$ are lepton universal 
by definition, and lead to lepton flavour non-universality only through lepton mass effects, which is not enough to provide explanation 
of the $R_{K^*}$ anomalies once bound from $B_d \to X_s \gamma$ is taken into account \cite{DescotesGenon:2011yn}. So we neglect 
NP effect in these operators. For all the other operators, we write their Wilson coefficients as $C_i = C_i^{\rm SM} + \Delta C_i $ where 
$\Delta C_i$ corresponds to the shift in the Wilson coefficient from its SM value due to short distance NP. 

%%%%%%%%%%%%%%%%%%%%%%%%%%%%%%%%%%%%%%%%%%%%%%%%%%%%%%%%%%%%%%%%%%%%%%%%%%%%%%%
%\vspace*{-5mm}
\subsection{\rm \bf Tensor~operators}
%\vspace*{-2mm}
%%%%%%%%%%%%%%%%%%%%%%%%%%%%%%%%%%%%%%%%%%%%%%%%%%%%%%%%%%%%%%%%%%%%%%%%%%%%%%%

In this section, we study the effect of the two tensor operators, ${\cal O}_T$, ${\cal O}_{T5}$, on $R_K$ and $R_{K^*}$. 
In Eq.~\ref{eq:tmp1} - \ref{eq:tmp3} below we show numerical formulae for the various branching ratios (normalised to their SM 
predictions) as functions of $\Delta C_{T5}^\mu$ and $\Delta C_{T5}^e$:
{\small
\noindent
\begin{align}
\frac{{\cal B}_{K\ell\ell}^{\rm cen}}{{\cal B}_{K\ell\ell}^{\rm cen}|_{\rm SM}} \approx&   1+  0.02 \, [\Delta C_{T5}^\ell]^2  \label{eq:tmp1}\\
\frac{{\cal B}_{K^*ee(\mu\mu)}^{\rm low}}{{\cal B}_{K^*ee(\mu\mu)}^{\rm low}|_{\rm SM}} \approx&  1 - 0.00 (0.24) \, 
[\Delta C_{T5}^{e(\mu)}]  + 0.30 \, [\Delta C_{T5}^{e(\mu)}]^2 \label{eq:tmp2} \\
\frac{{\cal B}_{K^*ee(\mu\mu)}^{\rm cen}}{{\cal B}_{K^*ee(\mu\mu)}^{\rm cen}|_{\rm SM}} \approx& 1 + 0.00(0.06) \, [\Delta C_{T5}^{e(\mu)}]  
+ 0.53 \, [\Delta C_{T5}^{e(\mu)}]^2 \label{eq:tmp3}
\end{align}}

The full set of numerical formulae valid in the presence of all the operators are presented in Appendix~\ref{num-exp}. These formulae can be used 
to perform very quick analysis of models as the only required inputs in these formulae are the short distance Wilson coefficients.

\begin{figure}[ht!]
\begin{center}
\begin{tabular}{cc}
\includegraphics[scale=0.32]{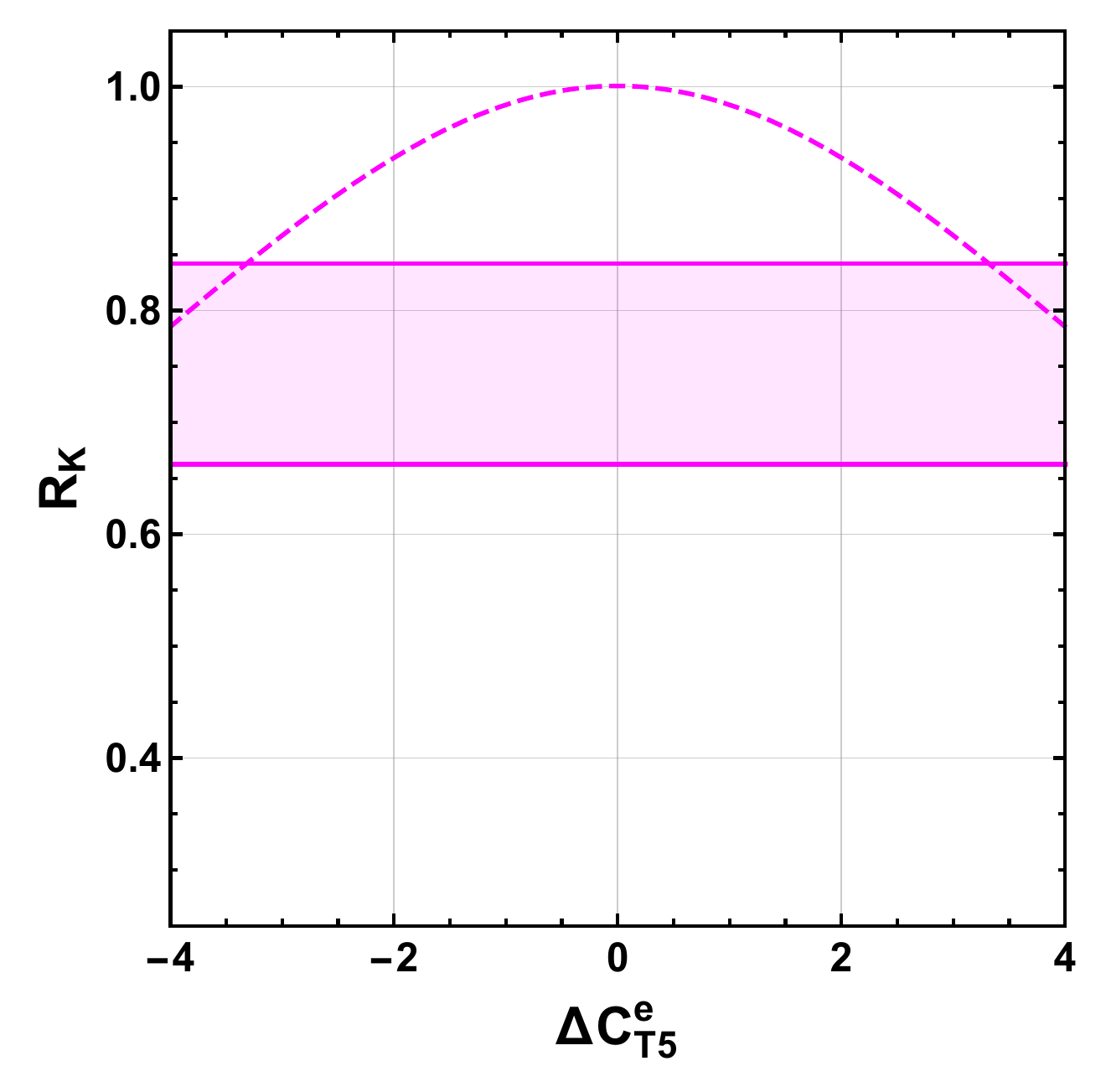} &
\hspace*{-3.5mm} \includegraphics[scale=0.32]{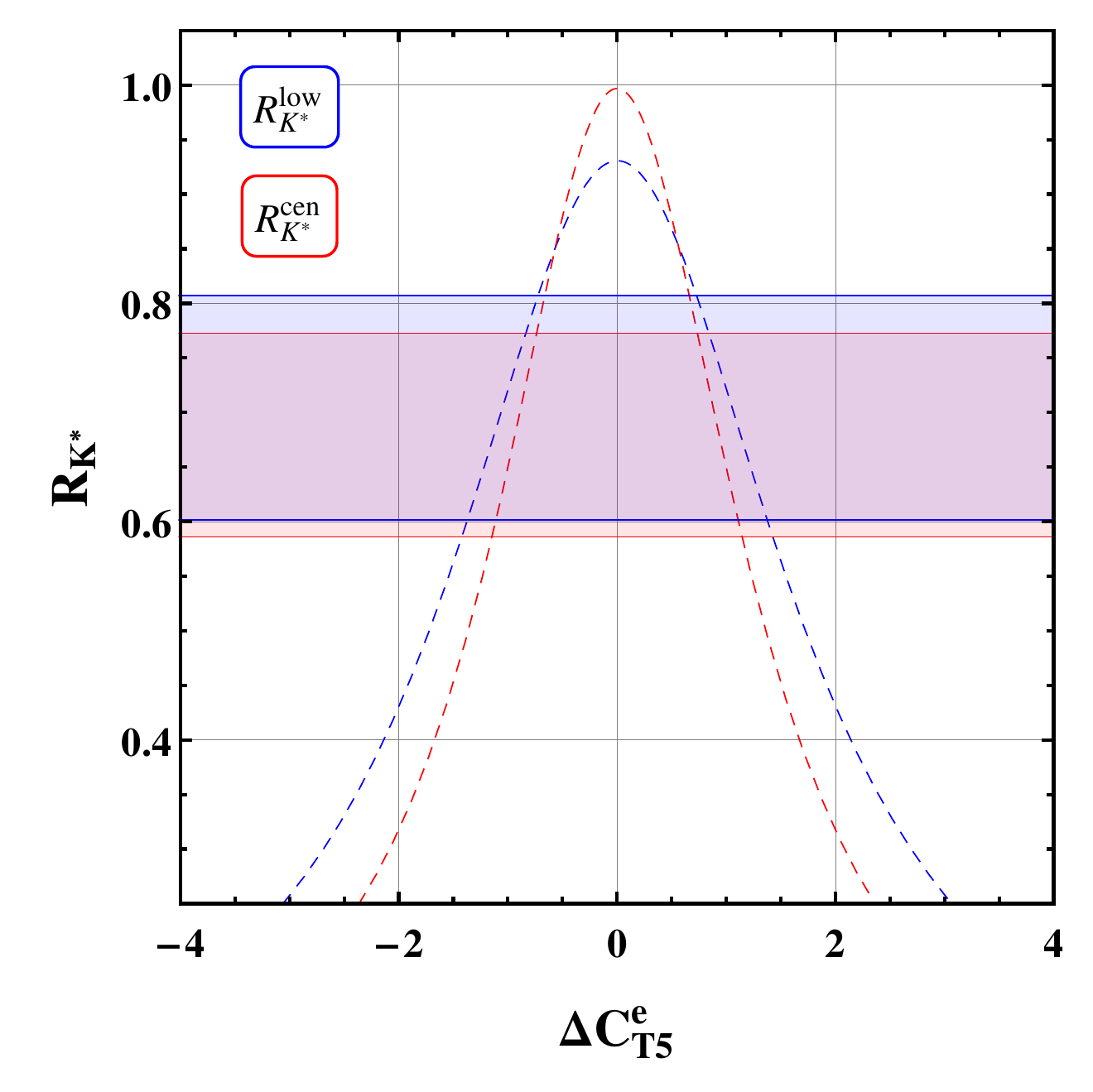}
\end{tabular}
\caption{Variation of $R_{K} ^{\rm cen}$, $R_{K^*} ^{\rm low}$ and $R_{K^*}^{\rm cen}$ with $\Delta C_{T5}^e$. The horizontal bands correspond to the 
experimental $1\sigma$ upper and lower limits shown in Table-\ref{tab:exp-data}. \label{fig:tensor}}
\end{center}
\end{figure}

In Fig.~\ref{fig:tensor} we show how $R_{K} ^{\rm cen}$, $R_{K^*} ^{\rm low}$ and $R_{K^*}^{\rm cen}$ vary with 
$\Delta C_{T5}^e$. 

%%%%%%%%%%%%%%%%%%%%%%%%%%%%%%%%%%%%%%%%%%%%%%%%%%%%%%%%%%%%%%%
%%%%%%%%%%%%%%%%%%%%%%%%%%%%%%%%%%%%%%%%%%%%%%%%%%%%%%%%%%%%%%%

It can be seen from the left panel of Fig.~\ref{fig:tensor} that $\Delta C_{T5}^e \sim \pm 1$ not only explains $R_{K^*}^{\rm cen}$ and 
$R_{K^*}^{\rm low}$ simultaneously but also brings them close to the experimental central values. As pointed out by one of the 
authors in \cite{Ghosh:2017ber}, this is not possible naturally by any other local operator at the dimension-6 level, and in this sense, the tensor 
operators are unique.  
However, as can be seen from the right panel of Fig.~\ref{fig:tensor}, $\Delta C_{T5}^e \sim \pm 1$ can not reduce $R_K^{\rm cen}$ much 
from its SM value of unity\footnote{That the tensor operators alone can not explain $R_K$ was also pointed out in \cite{Hiller:2014yaa}.}, and hence a simultaneous explanation of $R_K^{\rm cen}$, $R_{K^*}^{\rm cen}$ and $R_{K^*}^{\rm low}$ 
is not possible. All statements made here for $\Delta C_{T5}^{e}$ applies equally for the other tensor Wilson coefficient  $\Delta C_{T}^{e}$. 

Note that, any non-zero value for $\Delta C_{T}^\mu$ and $\Delta C_{T5}^\mu$ leads to values for $R_K$ and $R_{K^*}$ greater than their 
SM values and thus, tensor operators in the muon sector are ruled out as possible explanation of these anomalies. 

In the following section, we will investigate whether 
a simultaneous solution is possible when other additional operators are also considered. While we consider
only unprimed operators in the main text, the effect of the primed operators in conjunction with the
tensor operators can be found in Appendix \ref{sec:primed}. 

%%%%%%%%%%%%%%%%%%%%%%%%%%%%%%%%%%%%%%%%%%%%%%%%%%%%%%%%%%%%%%%%%%%%%%%%%%%%%%%
%\vspace*{-5mm}
\subsection{\text{\bf Combination of Vector and Tensor operators}}
\label{combination}
\begin{figure}[ht!]
\begin{center}
\begin{tabular}{cc}
\includegraphics[scale=0.45]{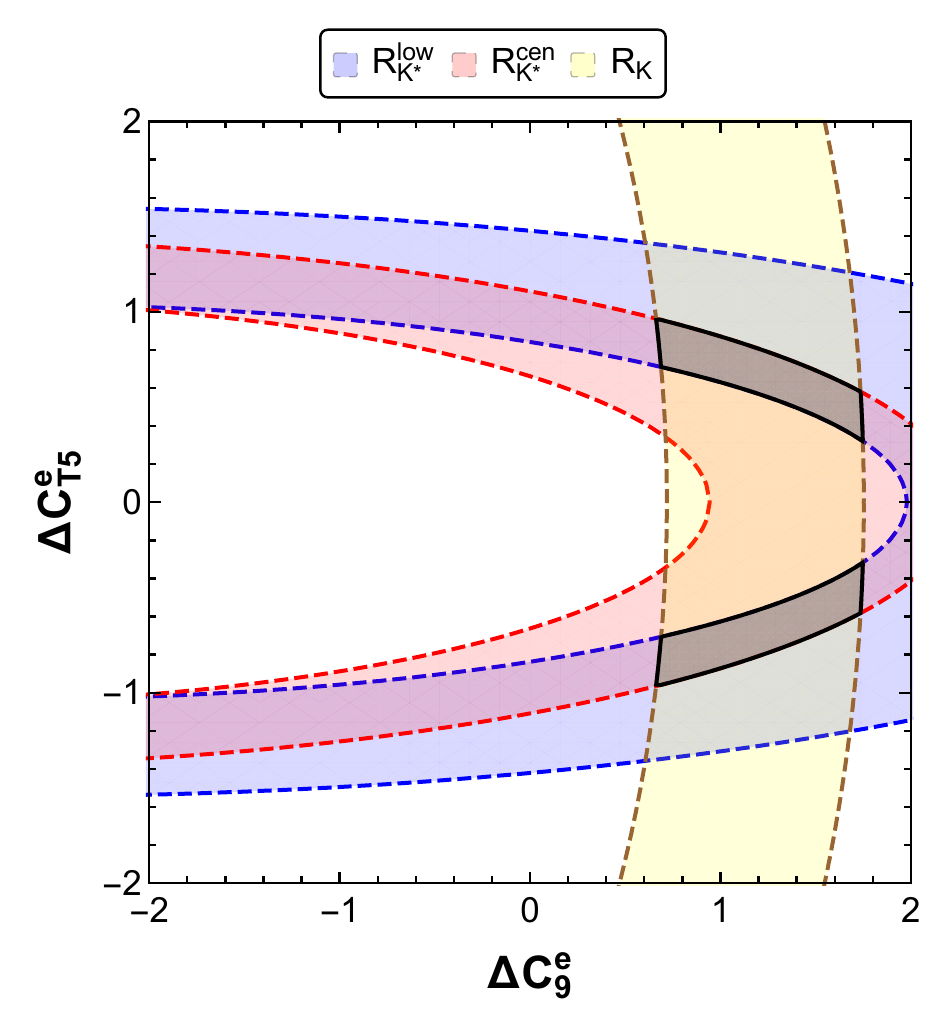} & 
\includegraphics[scale=0.45]{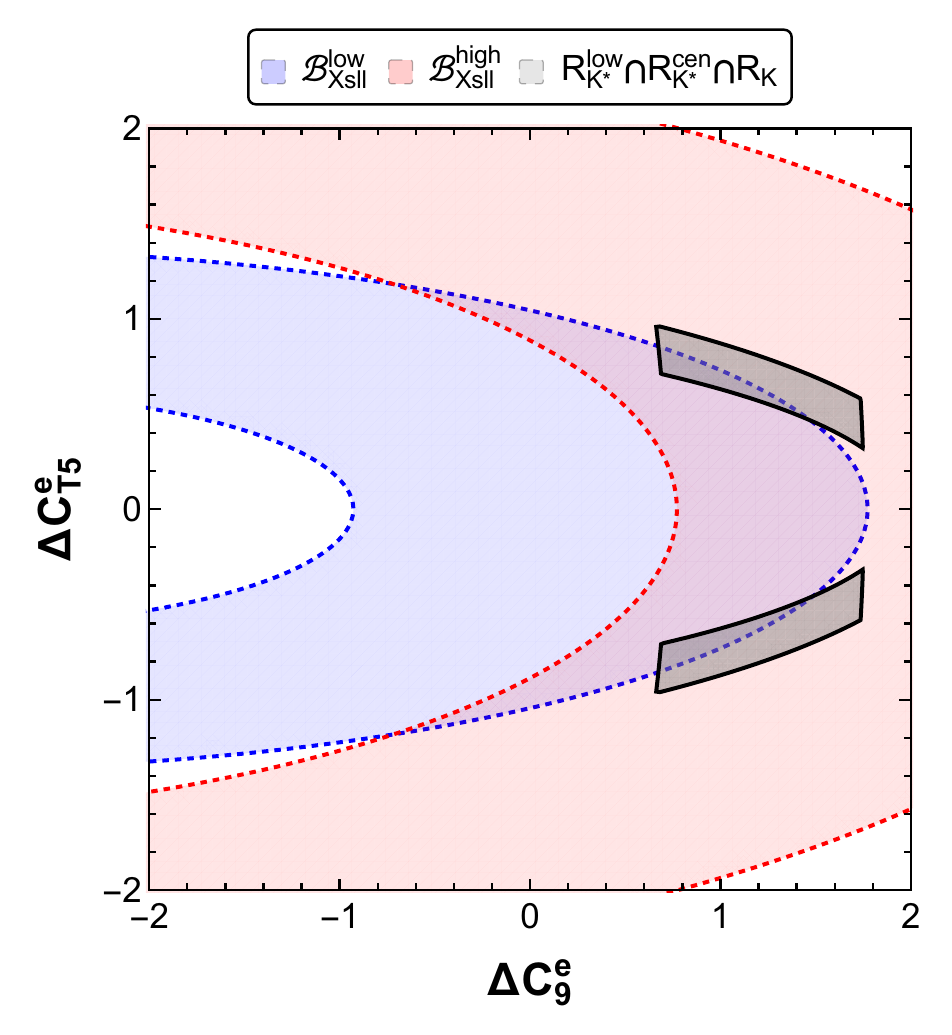}
\end{tabular}
\caption{Allowed regions in $\Delta C_9^e$ - $\Delta C_{T5}^e$ plane. See text for more details. \label{fig:c9ecT5e}}
\end{center}
\end{figure}

%\vspace*{-2mm}
%%%%%%%%%%%%%%%%%%%%%%%%%%%%%%%%%%%%%%%%%%%%%%%%%%%%%%%%%%%%%%%%%%%%%%%%%%%%%%%
In Fig.~\ref{fig:c9ecT5e}, we show the regions in $\Delta C_9^e$ - $\Delta C_{T5}^e$ plane allowed by the experimental measurements 
of the various observables listed in Table-\ref{tab:exp-data}. In the left panel, the blue, red and yellow shaded regions correspond to 
the $1\sigma$ experimental ranges of $R_{K^*}^{\rm low}$, $R_{K^*}^{\rm cen}$ and $R_K^{\rm cen}$ respectively. The black shaded regions 
are the overlap of the three. It should be noticed that the black shaded region is outside the $\Delta C_{T5}^e = 0$ line, and hence no 
simultaneous solutions are possible with only  $\Delta C_9^e$. In the right panel, we also show the regions allowed by 
$\mathcal{B}^{\text{low}}_{X_see}$(in blue) and $\mathcal{B}^{\text{high}}_{X_see}$ (in red). The black shaded region from the left panel 
is also superimposed there. It can be seen that there is a small overlap of the black, blue and red regions in the right panel where all the 
constraints including those from the inclusive decay are satisfied.

%%%%%%%%%%%%%%%%%%%%%%%%%%%%%%%%%%%%%%%%%%%%%%%%%%%%%%%%%%%%%%%
%%%%%%%%%%%%%%%%%%%%%%%%%%%%%%%%%%%%%%%%%%%%%%%%%%%%%%%%%%%%%%%

\begin{figure}[ht!]
\begin{center}
\begin{tabular}{cc}
\includegraphics[scale=0.45]{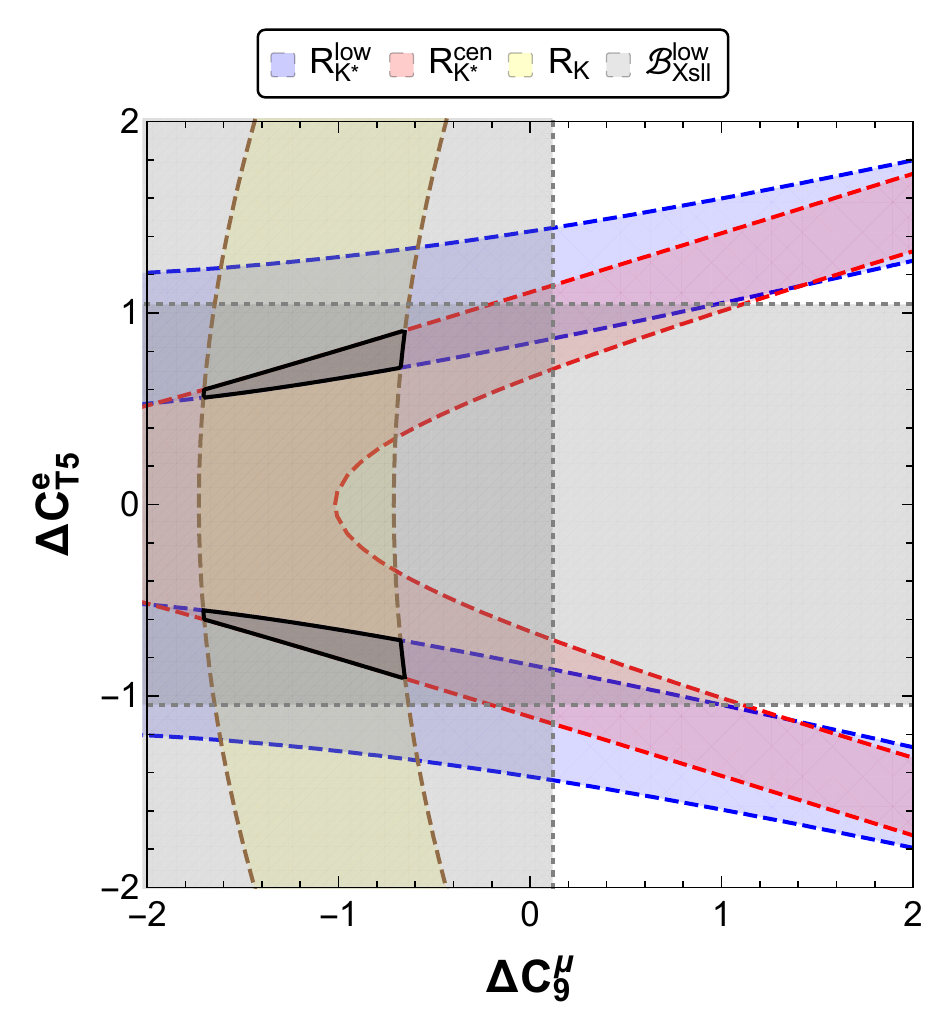} & 
\includegraphics[scale=0.45]{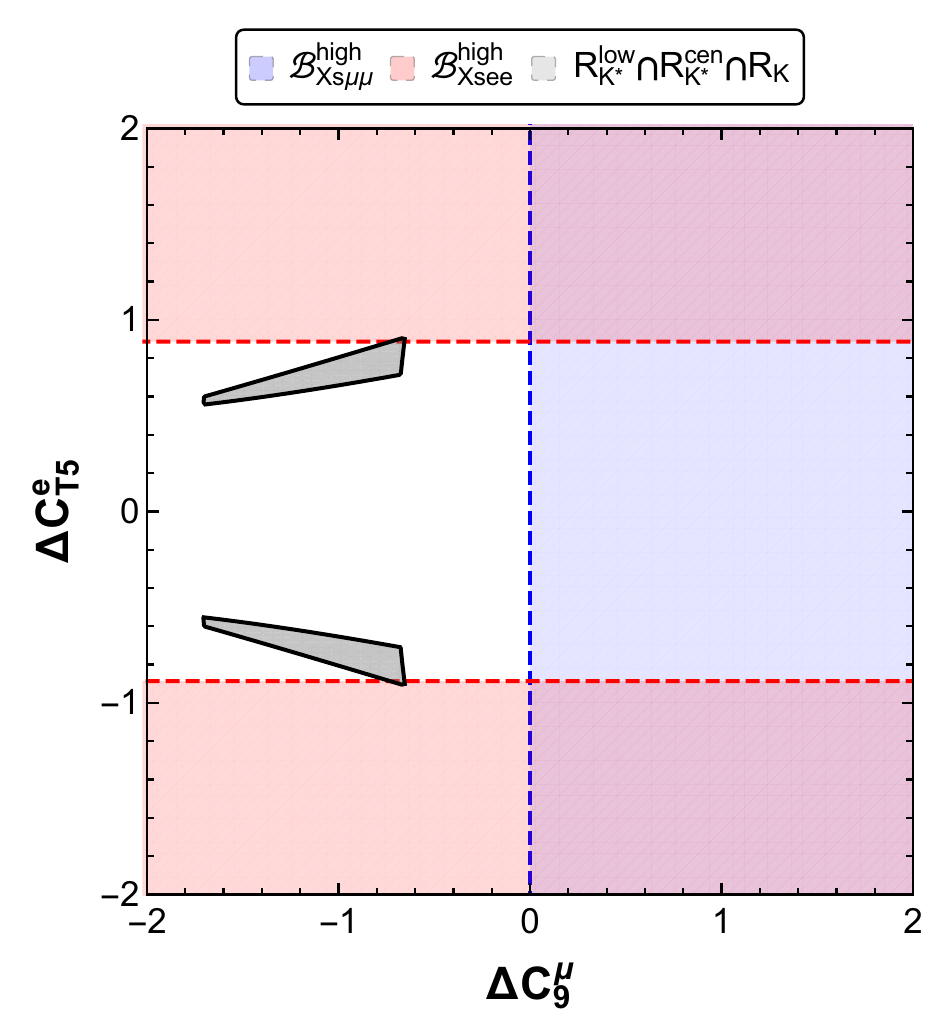}
\end{tabular}
\caption{Allowed regions in $\Delta C_9^\mu$ - $\Delta C_{T5}^e$ plane. See text for more details. \label{fig:c9mucT5e}}
\end{center}
\end{figure}

In Fig.~\ref{fig:c9mucT5e}, we show the allowed regions in the $\Delta C_9^\mu$ - $\Delta C_{T5}^e$ plane. The various shaded 
regions in the left panel have the same meaning as in Fig.~\ref{fig:c9ecT5e}. The grey vertical (horizontal) band corresponds to 
the experimental $1\sigma$ allowed region of $\mathcal{B}^{\text{low}}_{X_s\mu\mu}$ ($\mathcal{B}^{\text{low}}_{X_see}$). 
Similar to the previous case, here also a simultaneous solution is not possible with only $\Delta C_9^\mu$, and non-zero tensor contribution 
is required. However, as can be seen from the right panel of Fig.~\ref{fig:c9mucT5e}, this scenario is in tension with the measurements 
of  $\mathcal{B}^{\text{high}}_{X_s\ell\ell}$. 

%%%%%%%%%%%%%%%%%%%%%%%%%%%%%%%%%%%%%%%%%%%%%%%%%%%%%%%%%%%%%%%
%%%%%%%%%%%%%%%%%%%%%%%%%%%%%%%%%%%%%%%%%%%%%%%%%%%%%%%%%%%%%%%

\begin{figure}[ht!]
\begin{center}
\begin{tabular}{cc}
\includegraphics[scale=0.45]{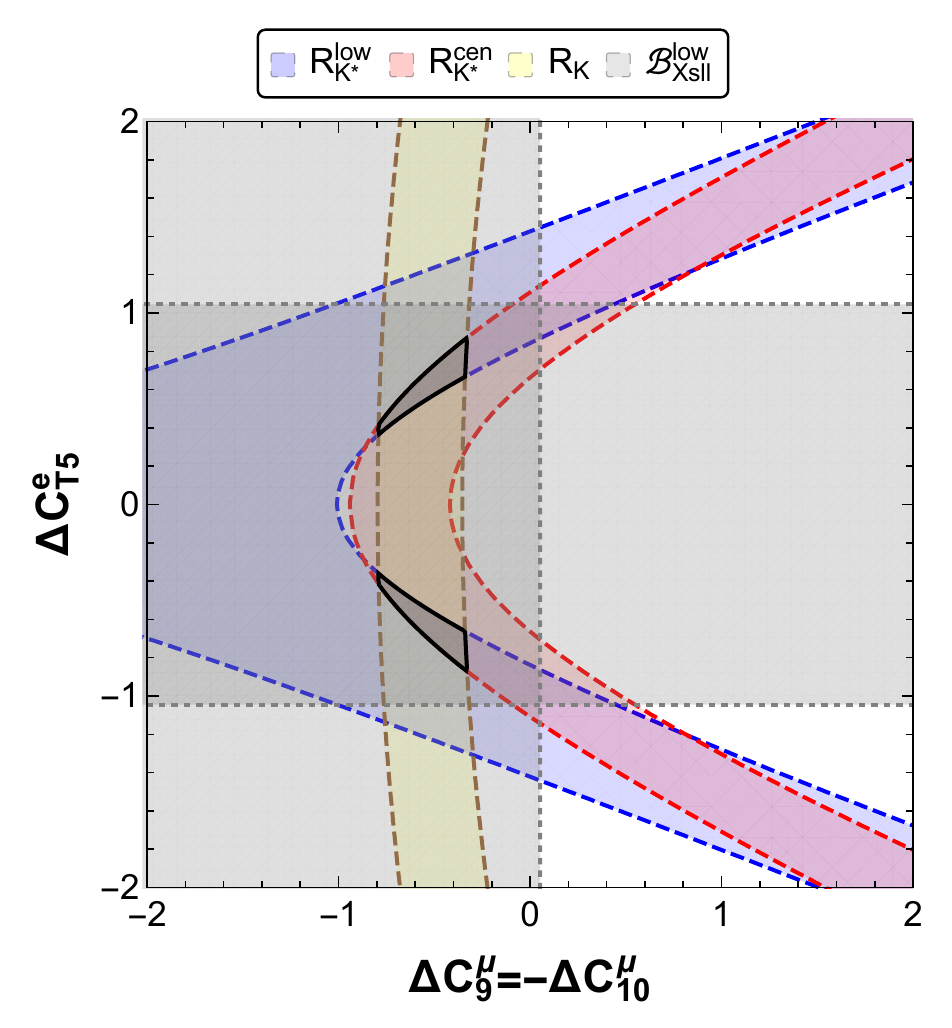} & 
\includegraphics[scale=0.45]{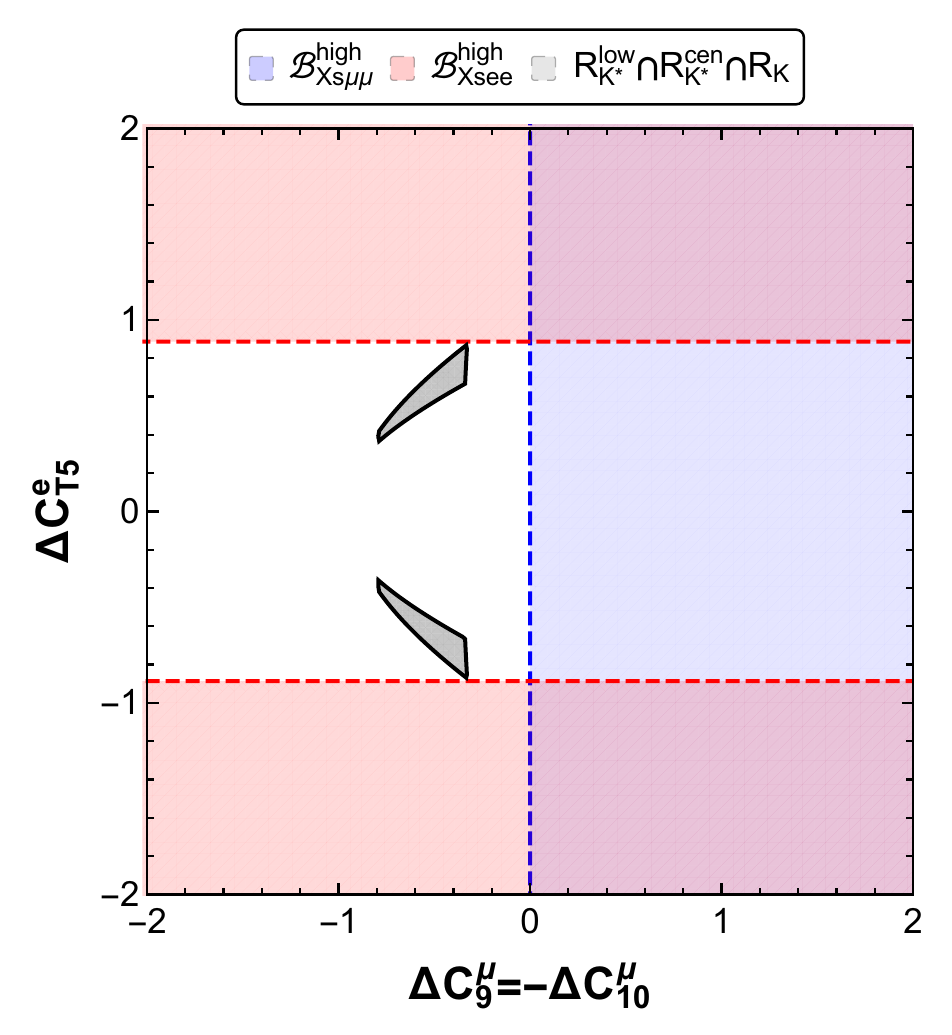} \\
\includegraphics[scale=0.45]{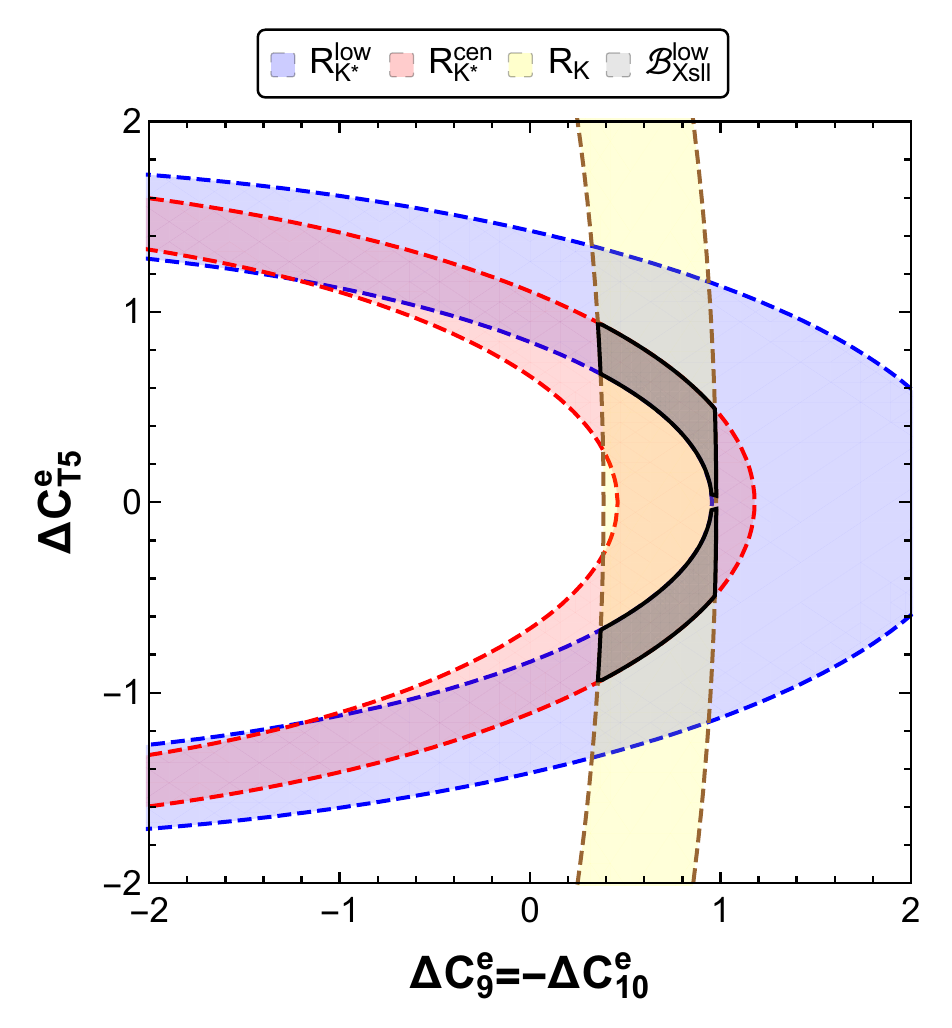} & 
\includegraphics[scale=0.45]{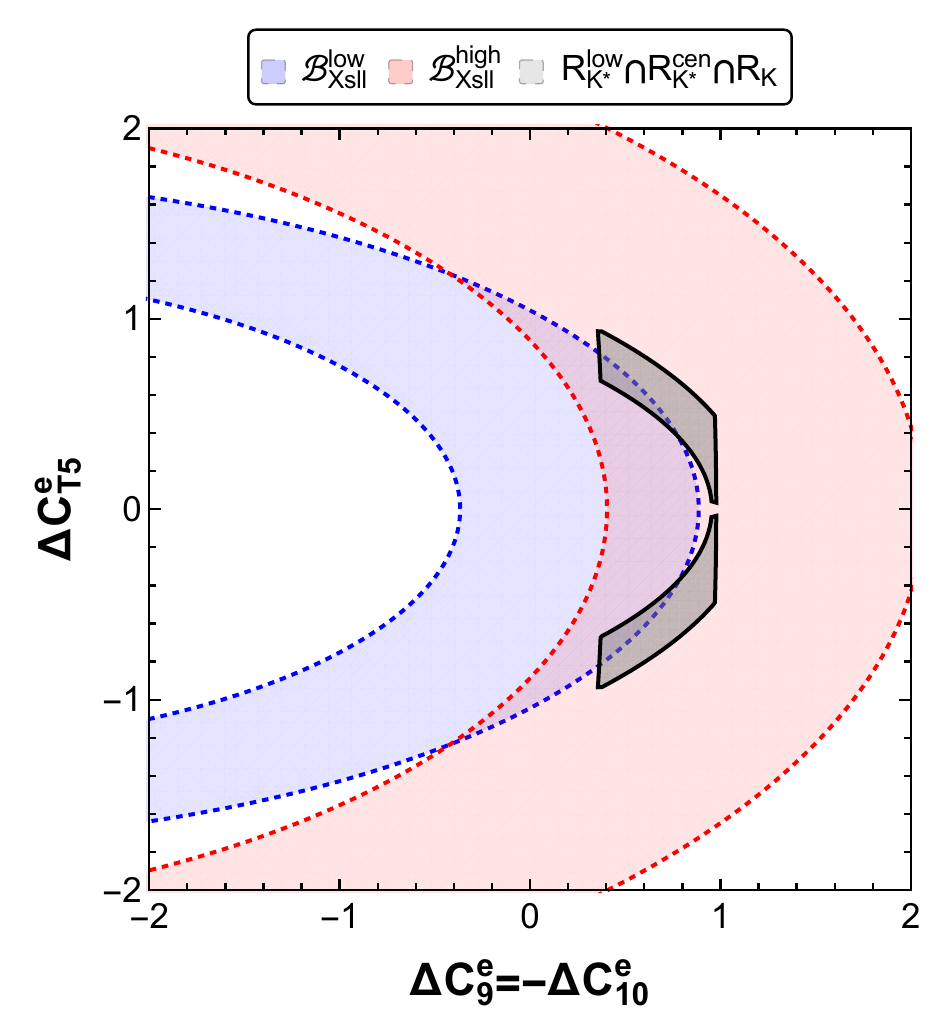}
\end{tabular}
\caption{Allowed regions in $\Delta C_9^\mu( = -\Delta C_{10}^\mu)$ - $\Delta C_{T5}^e$ plane (upper panel) and 
$\Delta C_9^e( = -\Delta C_{10}^e)$ - $\Delta C_{T5}^e$ plane (lower panel). See text for more details.
\label{fig:c9c10cT5}}
\end{center}
\end{figure}

We now consider the two cases $\Delta C_9^e = -\Delta C_{10}^e$ vs. $\Delta C_{T5}^e$ and 
$\Delta C_9^\mu = -\Delta C_{10}^\mu$ vs. $\Delta C_{T5}^e$. 
In Fig.~\ref{fig:c9c10cT5} we show our results. It can be seen from the upper panel that 
$\Delta C_9^\mu = -\Delta C_{10}^\mu$ alone (i.e., with $\Delta C_{T5}^e = 0$) can not explain $R_{K^*}^{\rm low}$, 
$R_{K^*}^{\rm cen}$ and $R_K^{\rm cen}$ simultaneously within their experimental $1\sigma$ regions. However, a simultaneous 
solutions is possible if a non-zero $\Delta C_{T5}^e \sim \pm 0.6$  is considered. 
Note that, the Wilson coefficient $C_{10}^\mu$ also modifies $\mathcal{B}_{\mu\mu}$ which gives a bound 
$0 \lesssim \Delta C_{10}^\mu \lesssim 0.7$ at the $1\sigma$ level \cite{Ghosh:2017ber}. Hence, the black overlap region in the upper left panel 
is allowed by $\mathcal{B}_{\mu\mu}$.
However, as in Fig.~\ref{fig:c9mucT5e}, this scenario also is in tension with the measurements of $\mathcal{B}^{\text{high}}_{X_s\ell\ell}$.

%%%%%%%%%%%%%%%%%%%%%%%%%%%%%%%%%%%%%%%%%%%%%%%%%%%%%%%%%%%%%%%
%%%%%%%%%%%%%%%%%%%%%%%%%%%%%%%%%%%%%%%%%%%%%%%%%%%%%%%%%%%%%%%%%%%%%%%%%%%%%%%

The situation is better for $\Delta C_9^e = -\Delta C_{10}^e$ vs. $\Delta C_{T5}^e$ as shown in the bottom panel of 
Fig.~\ref{fig:c9c10cT5}. Here, a simultaneous solutions to not only $R_{K^*}^{\rm low}$, 
$R_{K^*}^{\rm cen}$ and $R_K^{\rm cen}$, but also the inclusive decay ${\bar B}_d \to X_s e^+ e^-$ is possible. This 
corresponds to the small overlap of the black, red and blue shaded regions in the lower right panel of 
Fig.~\ref{fig:c9c10cT5}.

Before closing this section, we would like to mention that the tensor operators do not get generated at the dimension-6 level if 
$\rm SU(2) \times \rm U(1)_Y$ gauge invariance is imposed, which was also pointed out in \cite{Alonso:2014csa}. However, it can be generated at the dimension-8 level. For example, 
one can write down the operator $(1/\Lambda^4)(\overline{s_R} L_1 \tilde{H})\, (\overline{\mu_R} Q_3 \tilde{H})$ which, after electroweak 
symmetry breaking, generates the operator $(v^2/2\Lambda^4) (\bar s P_L \mu)\, (\bar \mu P_L b)$. This operator can be Fierz transformed 
into $(v^2/2\Lambda^4) (\bar s P_L b)\, (\bar \mu P_L \mu)$ and the tensor operator $(v^2/8\Lambda^4) (\bar s \sigma_{\mu\nu} P_L b)\, 
(\bar \mu \sigma^{\mu\nu} P_L \mu)$.  For more details, see Appendix~\ref{su2u1}.

%%%%%%%%%%%%%%%%%%%%%%%%%%%%%%%%%%%%%%%%%%%%%%%%%%%%%%%%%%%%%%%%%%%%%%%%%%%%%%%
%\vspace*{-5mm}
\subsection{\text{\bf Summary}}
%\vspace*{-2mm}
%%%%%%%%%%%%%%%%%%%%%%%%%%%%%%%%%%%%%%%%%%%%%%%%%%%%%%%%%%%%%%%%%%%%%%%%%%%%%%%
Motivated by the recent measurements of $R_{K^*}$ in two $q^2$ bins by the LHCb collaboration, we have performed a detailed analysis of the role 
of tensor operators in $R_K$ and $R_{K^{*}}$, for the first time in the literature. We show that, unlike the vector, axial vector, scalar or pseudo scalar operators, tensor 
operators can comfortably explain $R_{K^*}^{\rm cen}$ and $R_{K^*}^{\rm low}$ simultaneously. Hence, if the experimental measurement of 
$R_{K^*}$ in the low $q^2$ bin stays in the future, either a very light vector boson (as shown by one of the authors in \cite{Ghosh:2014awa}) or 
the existence of tensor operators would be unavoidable. However, we find that a simultaneous explanation of $R_K$ 
also would require the existence of other Wilson coefficients (of vector and/or axial vector operators, for example) in conjunction with the tensor operators. 
We study the interplay of the vector and axial vector operators with the tensor structures, and obtain the regions allowed by the $1\sigma$ experimental 
values of $R_K$ and $R_{K^*}$. 
We further show that the measured branching ratios for the inclusive $B_d \to X_s \ell^+ \ell^-$ decay provide very important constraints on the various solutions. We also present completely general numerical formulae which can be used to effortlessly compute $R_{K}^{\rm cen}$, $R_{K^*}^{\rm cen}$, $R_{K^*}^{\rm low}$ and the inclusive branching fractions just knowing the short distance Wilson coefficients at the $m_b$ scale.  
%%%%%%%%%%%%%%%%%%%%%%%%%%%%%%%%%%%%%%%%%%%%%%%%%%%%%%%%%%%%%%%%%%%%%%%%%%%%%
\onecolumngrid
\vspace*{2mm}
\begin{center}
{\bf 
--------------------------------------------------------------------------------
}
\end{center}

\appendix
\label{appendix}
%%%%%%%%%%%%%%%%%%%%%%%%%%%%%%%%%%%%%%%%%%%%%%%%%%%%%%%%%%%%%%%%%%%%%%%%%%%%%%%
\onecolumngrid
\vspace{-1cm}
\section{Complete expressions for the branching ratios}
\label{num-exp}
\vspace{-5mm}

\bea
\dfrac{{\cal B}_{Kee}^{\rm cen}}{{\cal B}_{Kee}^{\rm cen}|_{\rm SM}} 
=   
1 &&+0.2429 \cNINEe+0.0274 \cNINEe^2+0.2429 \cNINEpe+0.0549 \cNINEe 
\cNINEpe+0.0274 \cNINEpe^2- 0.225 \cOTENe \nn\\ [-2mm]
&& +0.0274 \cOTENe^2-0.225 \cOTENpe+0.0549 \cOTENe \cOTENpe+0.0274 
\cOTENpe^2+0.0092 \cSe^2\nn\\
&& +0.0184 \cSe \cSpe+0.0092 \cSpe^2+0.0092 \cssPe^2
+0.0184 \cssPe \cssPpe+0.0092 \cssPpe^2\nn\\
&&+0.0002 \cTe+0.0171 \cTe^2+0.0171 \cTFIVEe^2 \\ [2mm]
% \eea
% 
% \bea
\dfrac{{\cal B}_{K\mu\mu}^{\rm cen}}{{\cal B}_{K\mu\mu}^{\rm 
cen}|_{\rm SM}} = 1 &&+0.2427 \cNINEmu+0.0274 \cNINEmu^2+0.2427 
\cNINEpmu+0.0548 \cNINEmu \cNINEpmu+0.0274 \cNINEpmu^2-0.2253 \cOTENmu\nn\\ 
[-2mm]
&&+0.0275 \cOTENmu^2-0.225 \cOTENpmu+0.055 \cOTENmu \cOTENpmu+0.0275\cOTENpmu^2 
+ 0.009 \cSmu^2\nn\\
&&+0.018 \cSmu  \cSpmu +0.009 \cSpmu^2-0.0187 \cssPmu + 0.0046 \big( \cOTENmu + 
\cOTENpmu \big)\big( \cssPmu + \cssPpmu\big)\nn\\
&&+0.0091 \cssPmu^2-0.0187 \cssPpmu+0.0182 \cssPmu \cssPpmu+0.0091 
\cssPpmu^2+0.0168 \cTFIVEmu^2\nn\\
&&+0.0457 \cTmu+0.0103\big( \cNINEmu + 
\cNINEpmu\big) \cTmu+0.0174 \cTmu^2 \\ [2mm]
% \eea
% 
% \bea
\dfrac{ {\cal B}_{K^*ee}^{\rm low}}{ {\cal B}_{K^*ee}^{\rm low}|_{\rm SM}} 
= 1 && +0.0764 \cNINEe+0.0136 \cNINEe^2-0.1048 \cNINEpe-0.0257 \cNINEe 
\cNINEpe+0.0136 \cNINEpe^2-0.1118 \cOTENe\nn\\ [-2mm]
&&+0.0136 \cOTENe^2+0.1054 \cOTENpe-0.0257 \cOTENe \cOTENpe+0.0136 
\cOTENpe^2+0.0006\big( \cSe -\cSpe \big)^2\nn\\
&&+0.0006 \big( \cssPe -\cssPpe \big)^2 -0.0015 \cTe+0.2901 \cTe^2-0.0013 
\cTFIVEe+0.2901 \cTFIVEe^2\\[2mm]
% \eea
% 
% \bea
\dfrac{ {\cal B}_{K^*\mu\mu}^{\rm low}}{ {\cal B}_{K^*\mu\mu}^{\rm low}|_{\rm 
SM}} = 1 && + 0.0806 \cNINEmu+0.0144 \cNINEmu^2-0.1103 \cNINEpmu-0.027 \cNINEmu 
\cNINEpmu+0.0144 \cNINEpmu^2-0.1167 \cOTENmu\nn\\ [-2mm]
&& +0.0142 \cOTENmu^2+0.1106 \cOTENpmu-0.027 \cOTENmu \cOTENpmu+0.0142 
\cOTENpmu^2+0.0006 \cSmu^2\nn\\
&&-0.0012 \cSmu \cSpmu+0.0006 \cSpmu^2-0.0078 \cssPmu+0.0019 \cOTENmu 
\cssPmu-0.0019 \cOTENpmu \cssPmu \nn\\
&& + 0.0006 \cssPmu^2+0.0078 \cssPpmu-0.0019 \cOTENmu \cssPpmu+0.0019 \cOTENpmu 
\cssPpmu-0.0013 \cssPmu \cssPpmu\nn\\
&&+0.0006 \cssPpmu^2-0.2362 \cTFIVEmu+0.0165 \cNINEmu \cTFIVEmu-0.0165 
\cNINEpmu 
\cTFIVEmu+0.3057 \cTFIVEmu^2\nn\\
&&-0.2676 \cTmu+0.0088 \cNINEmu \cTmu+0.0088 \cNINEpmu \cTmu+0.305 \cTmu^2
%[2mm]
\eea
 \bea
\dfrac{ {\cal B}_{K^*ee}^{\rm cen}}{ {\cal B}_{K^*ee}^{\rm cen}|_{\rm 
SM}} = 1 &&+ 0.2187 \cNINEe+0.032 \cNINEe^2-0.1998 \cNINEpe-0.0474 \cNINEe 
\cNINEpe+0.032 \cNINEpe^2-0.2629 \cOTENe\nn\\ [-2mm]
&&+0.032 \cOTENe^2+0.1945 \cOTENpe-0.0474 \cOTENe \cOTENpe+0.032 
\cOTENpe^2+0.0067 \cSe^2\nn\\
&&-0.0134 \cSe \cSpe+0.0067 \cSpe^2+0.0067 \cssPe^2-0.0134 \cssPe 
\cssPpe+0.0067 \cssPpe^2\nn\\
&&+0.5349 \cTe^2+0.0003 \cTFIVEe+0.0001 \cNINEe \cTFIVEe-0.0001 \cNINEpe 
\cTFIVEe+0.5349 \cTFIVEe^2
%\\[0.0mm]
\eea

\bea
\dfrac{ {\cal B}_{K^*\mu\mu}^{\rm cen}}{ {\cal B}_{K^*\mu\mu}^{\rm 
cen}|_{\rm SM}} =  1 &&+ 0.2194 \cNINEmu+0.0321 \cNINEmu^2-0.2004 
\cNINEpmu-0.0476 \cNINEmu \cNINEpmu+0.0321 \cNINEpmu^2-0.2622 
\cOTENmu\nn\\ [-2mm]
&&+0.032 \cOTENmu^2+0.1949 \cOTENpmu-0.0475 \cOTENmu \cOTENpmu+0.032 
\cOTENpmu^2+0.0066 \cSmu^2\nn\\
&&-0.0132 \cSmu \cSpmu+0.0066 \cSpmu^2-0.0138 \cssPmu+0.0034 \cOTENmu 
\cssPmu-0.0034 \cOTENpmu \cssPmu\nn\\
&&+0.0067 \cssPmu^2+0.0138 \cssPpmu-0.0034 \cOTENmu \cssPpmu+0.0034 \cOTENpmu 
\cssPpmu-0.0134 \cssPmu \cssPpmu\nn\\
&&+0.0067 \cssPpmu^2+0.0638 \cTFIVEmu+0.0295 \cNINEmu \cTFIVEmu-0.0295 
\cNINEpmu 
\cTFIVEmu+0.5373 \cTFIVEmu^2\nn\\
&&+0.0039 \cTmu+0.0154 \cNINEmu \cTmu+0.0154 \cNINEpmu \cTmu+0.5359 \cTmu^2
 \eea
 \bea
10^6 {\cal B}_{X_s\ell\ell}^{\rm low} = && 10^6 {\cal B}_{X_s\ell\ell}^{\rm 
low}|_{\rm SM}   +0.4156 \cNINEell+0.0647 \big(
\cNINEell^2+ \cNINEpell^2+ \cOTENell^2+ \cOTENpell^2  \big) -0.5308 \cOTENell 
\nn\\
&&+0.0108 \big( \cSell^2+ \cSpell^2+ \cssPell^2+ \cssPpell^2\big) +0.8615 
\big( \cTFIVEell^2+  \cTell^2 \big) \\ [2mm]
10^6 {\cal B}_{X_s\ell\ell}^{\rm high} = && 10^6 {\cal B}_{X_s\ell\ell}^{\rm 
high}|_{\rm SM}   + 0.1187 \cNINEell+0.0143 
\big( \cNINEell^2+ \cNINEpell^2 + \cOTENell^2\big) -0.1171 \cOTENell  + 0.0143 
\cOTENpell^2 \nn\\
&& +0.0063 \big( \cSell^2+ \cSpell^2+ \cssPell^2+  \cssPpell^2 
\big) + 0.1272 \big( \cTFIVEell^2+ \cTell^2 \big)
\eea

\vspace*{15mm}
\twocolumngrid
%%%%%%%%%%%%%%%%%%%%%%%%%%%%%%%%%%%%%%%%%%%%%%%%%%%%%%%%%%%%%%%%%%%%%%%%%%%%%%%
\section{\text{\bf Primed operators}}
\label{sec:primed}
%\vspace*{-3mm}

Earlier we considered only the unprimed vector and axial vector 
operators namely, $C_9^{\mu, e}$ and $C_{10}^{\mu, e}$, and neglected 
their primed counterparts $C_{9^\prime}^{\mu, e}$ and $C_{10^\prime}^{\mu, e}$. 
It has been shown (see for example, \cite{Ghosh:2014awa}) that 
the primed operators alone are unable to produce the experimental measurements 
of $R_K$ and $R_{K^*}$ simultaneously. In this section, we 
will investigate whether the situation can improve in the presence of tensor 
operators. 

\begin{figure}[htb!]
\begin{center}
\begin{tabular}{cc}
\includegraphics[scale=0.45]{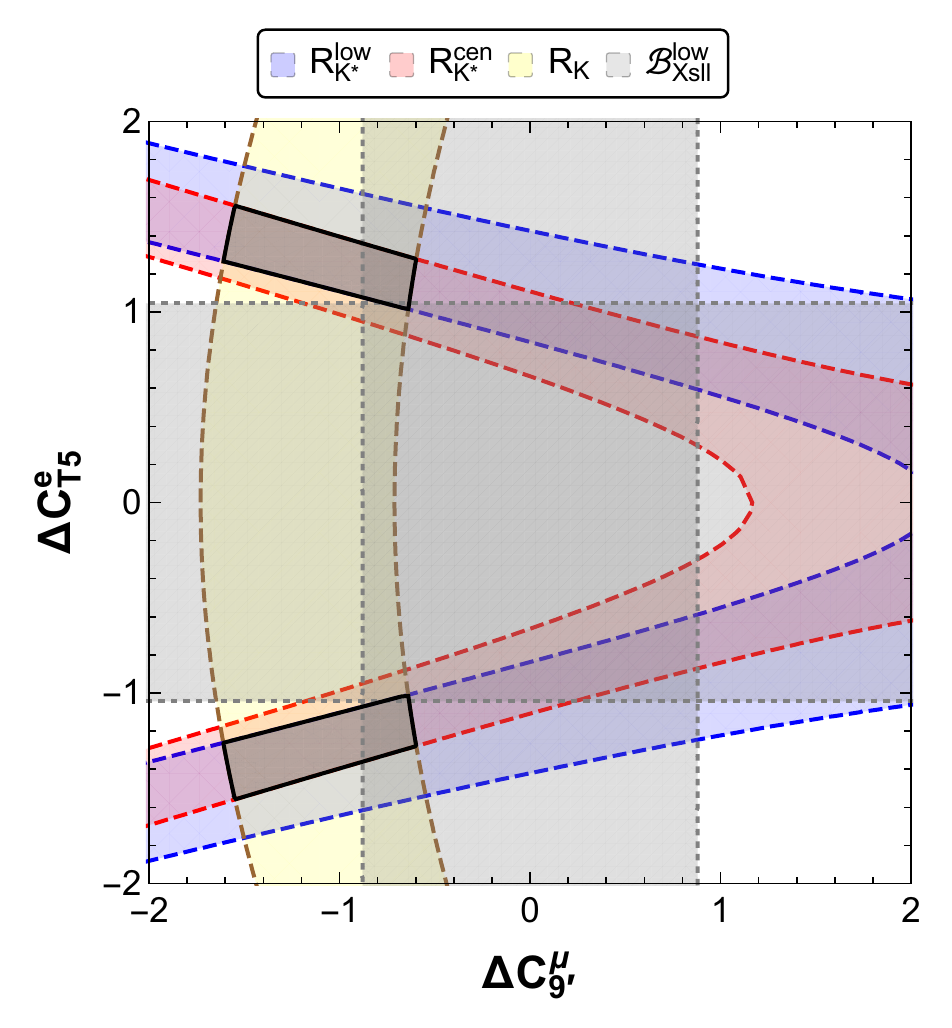} & 
\includegraphics[scale=0.45]{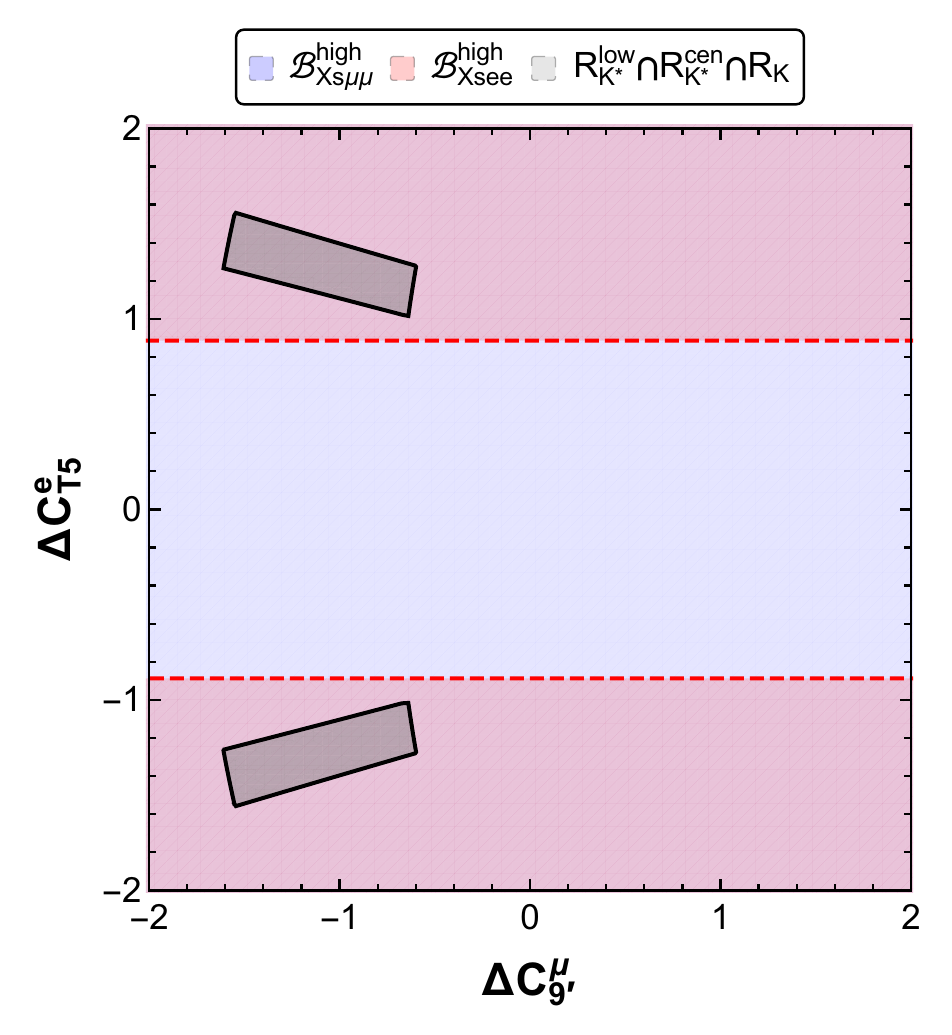}
\end{tabular}
\caption{Allowed regions in $\Delta C_{9'}^\mu$ - $\Delta C_{T5}^e$ plane. See 
text for more details.
\label{fig:c9pmucT5}}
\end{center}
\end{figure}

Fig.~\ref{fig:c9pmucT5} shows the allowed regions in $\Delta C_{9'}^\mu$ - 
$\Delta C_{T5}^e$ plane.

%%%%%%%%%%%%%%%%%%%%%%%%%%%%%%%%%%%%%%%%%%%%%%%%%%%%%%%%%%%%%%%
%%%%%%%%%%%%%%%%%%%%%%%%%%%%%%%%%%%%%%%%%%%%%%%%%%%%%%%%%%%%%%%

It can be seen that in order to satisfy $R_K^{\rm cen}$, $R_{K^\ast}^{\rm low}$ 
and $R_{K^\ast}^{\rm cen}$ simultaneously in the 
presence of $\Delta C_{9'}^\mu$, large value of $\Delta C_{T5}^e \approx \pm 
1.3$ is also needed.

%%%%%%%%%%%%%%%%%%%%%%%%%%%%%%%%%%%%%%%%%%%%%%%%%%%%%%%%%%%%%%%
\begin{figure}[htb!]
\begin{center}
\begin{tabular}{cc}
\includegraphics[scale=0.45]{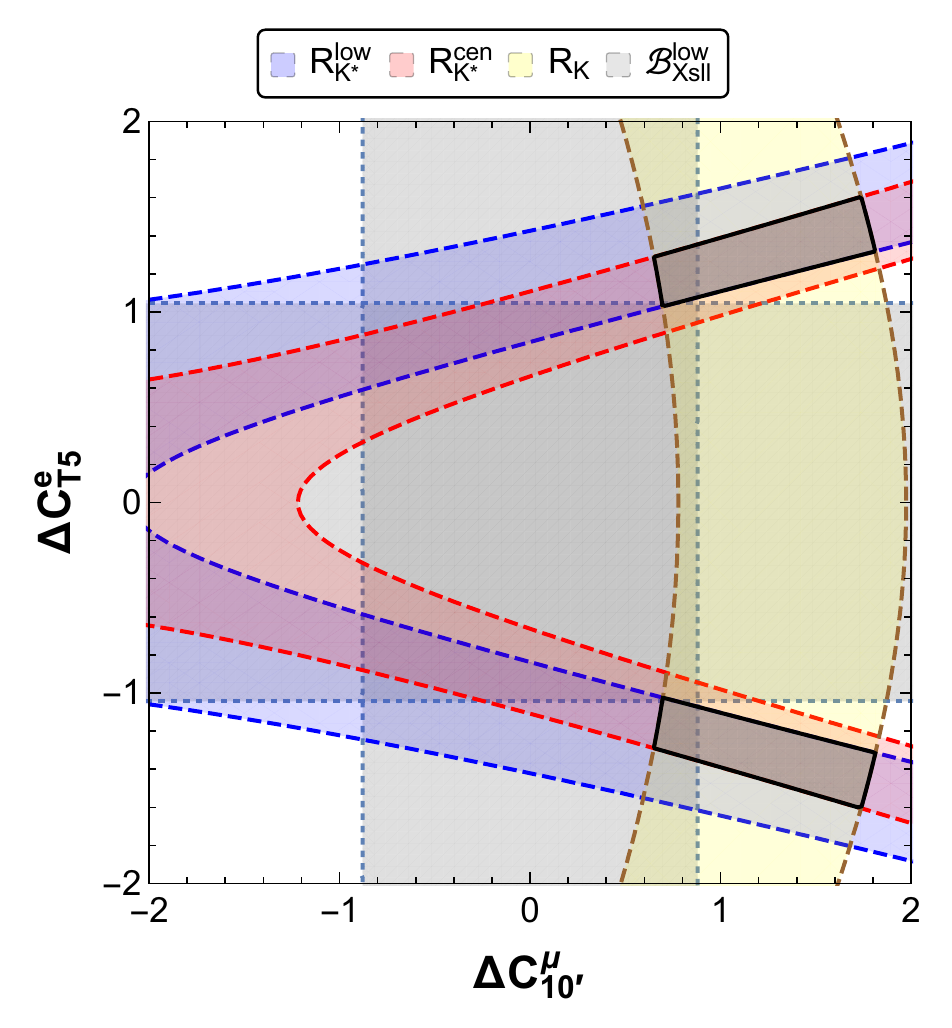} & 
\includegraphics[scale=0.45]{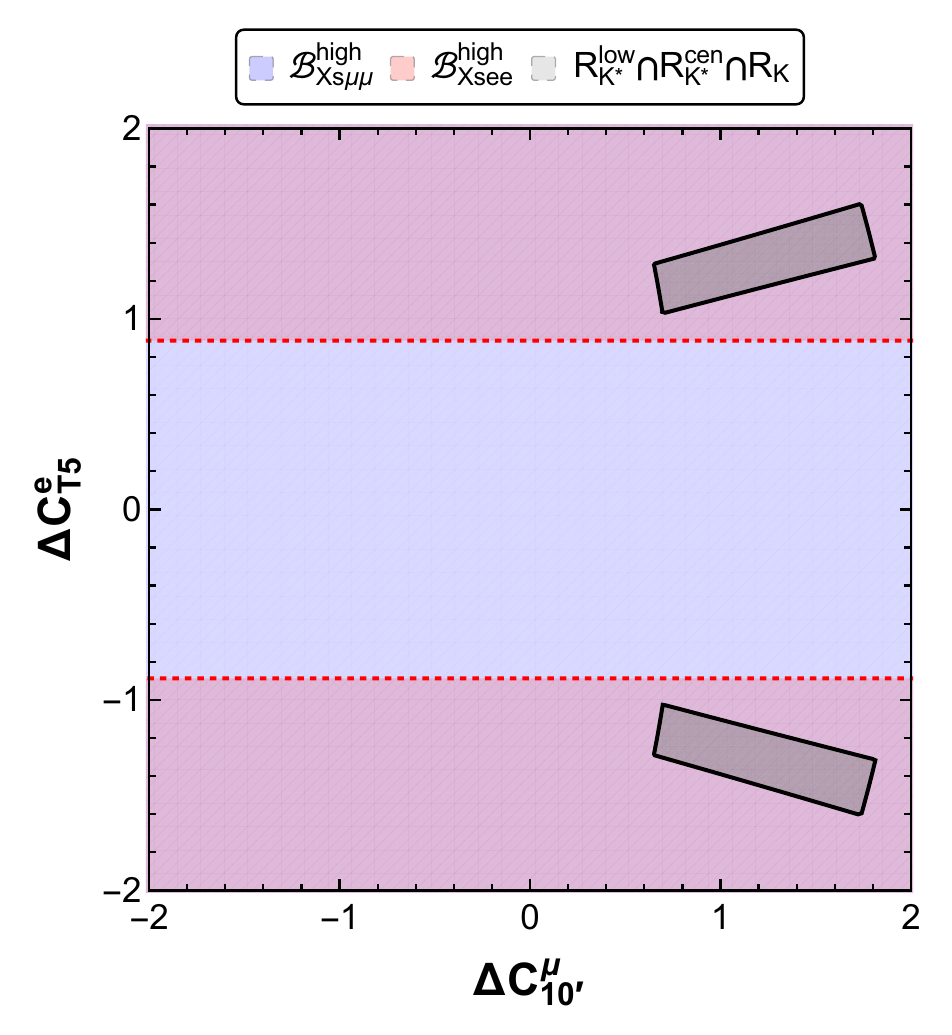}
\end{tabular}
\caption{Allowed regions in $\Delta C_{10'}^\mu$ - $\Delta C_{T5}^e$ plane. See 
text for more details.
\label{fig:c10pmucT5}}
\end{center}
\end{figure}
%%%%%%%%%%%%%%%%%%%%%%%%%%%%%%%%%%%%%%%%%%%%%%%%%%%%%%%%%%%%%%%

However, this solution is in tension with $\mathcal{B}^{\text{low}}_{X_see}$ as 
can be seen from the grey region in the left panel 
of Fig.~\ref{fig:c9pmucT5}. Note that, in the right panel of 
Fig.~\ref{fig:c9pmucT5} the blue region covers the whole plane, and hence 
this solution is consistent with $\mathcal{B}^{\text{high}}_{X_s\ell\ell}$.
Similar statements can be made also for $\Delta C_{10'}^\mu$, as can be seen 
from Fig.~\ref{fig:c10pmucT5}.

%%%%%%%%%%%%%%%%%%%%%%%%%%%%%%%%%%%%%%%%%%%%%%%%%%%%%%%%%%%%%%%
\begin{figure}[htb!]
\begin{center}
\begin{tabular}{cc}
\includegraphics[scale=0.45]{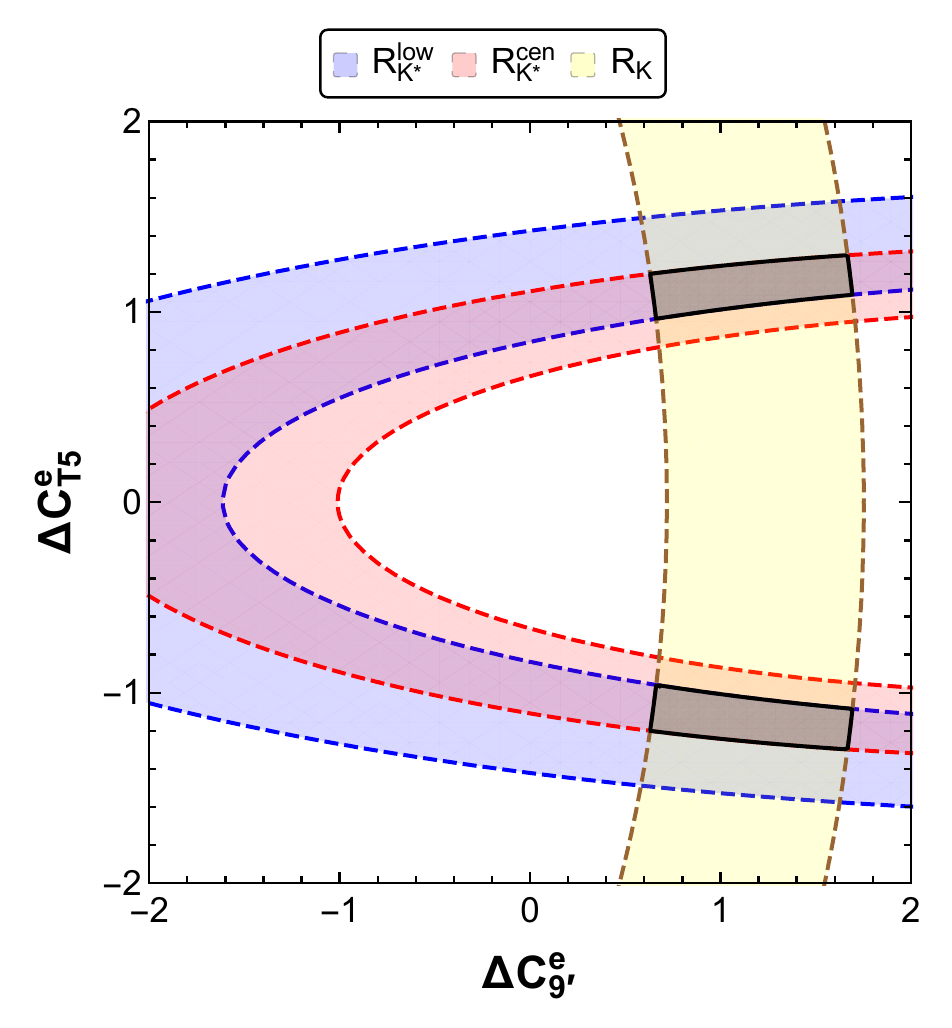} & 
\includegraphics[scale=0.45]{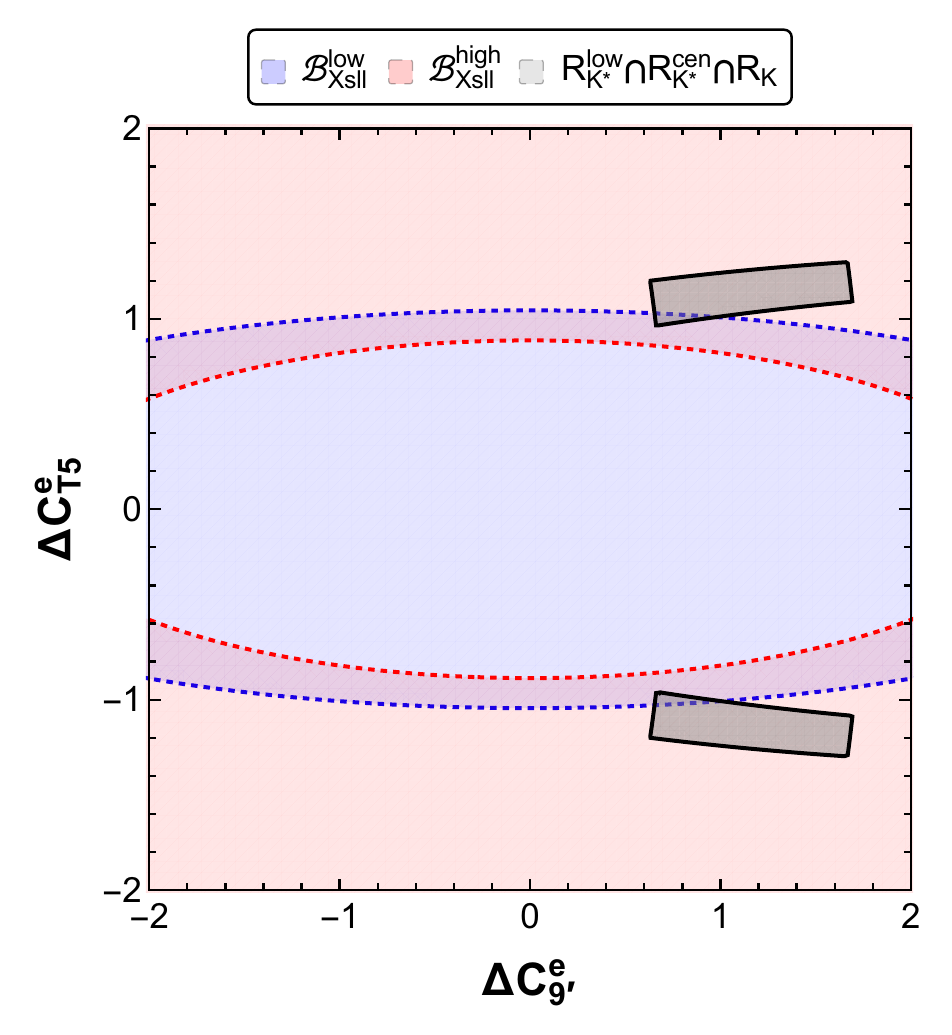}
\end{tabular}
\caption{Allowed regions in $\Delta C_{9'}^e$ - $\Delta C_{T5}^e$ plane. See 
text for more details.
\label{fig:c9pecT5}}
\end{center}
\end{figure}
%%%%%%%%%%%%%%%%%%%%%%%%%%%%%%%%%%%%%%%%%%%%%%%%%%%%%%%%%%%%%%%

%%%%%%%%%%%%%%%%%%%%%%%%%%%%%%%%%%%%%%%%%%%%%%%%%%%%%%%%%%%%%%%
\begin{figure}[htb!]
\begin{center}
\begin{tabular}{cc}
\includegraphics[scale=0.45]{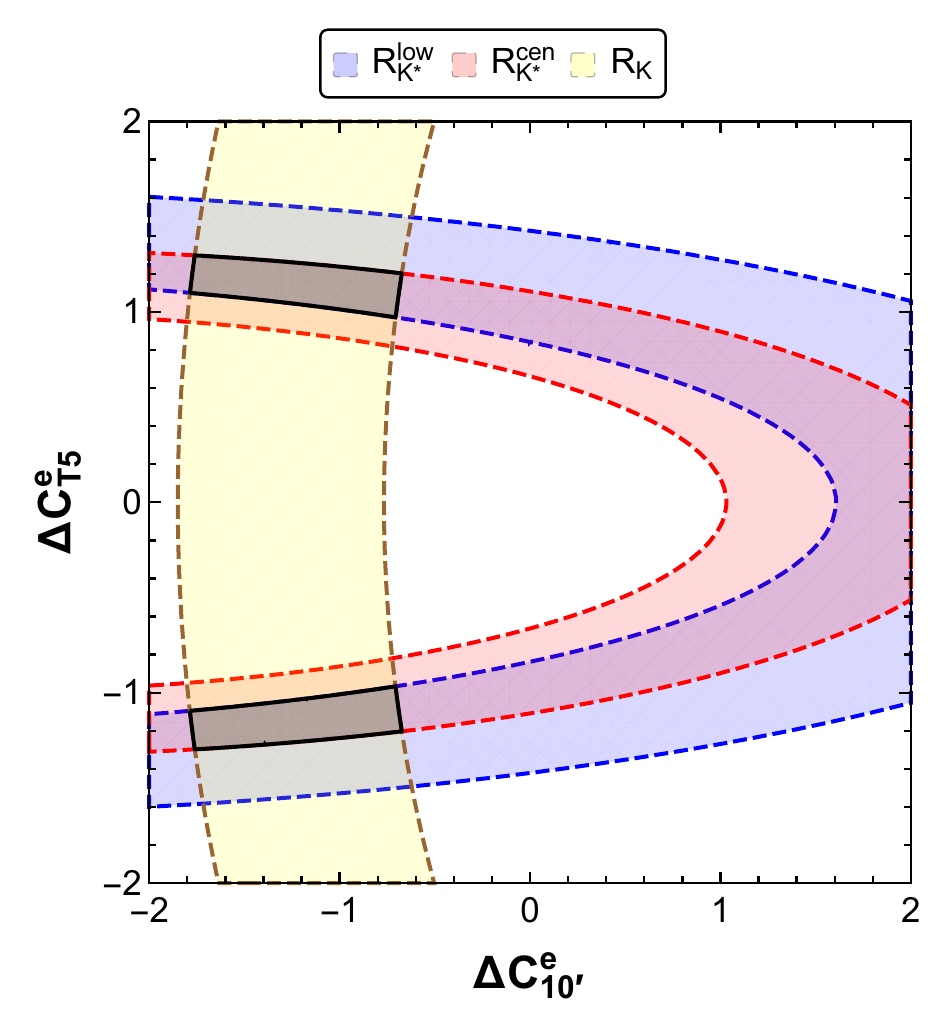} & 
\includegraphics[scale=0.45]{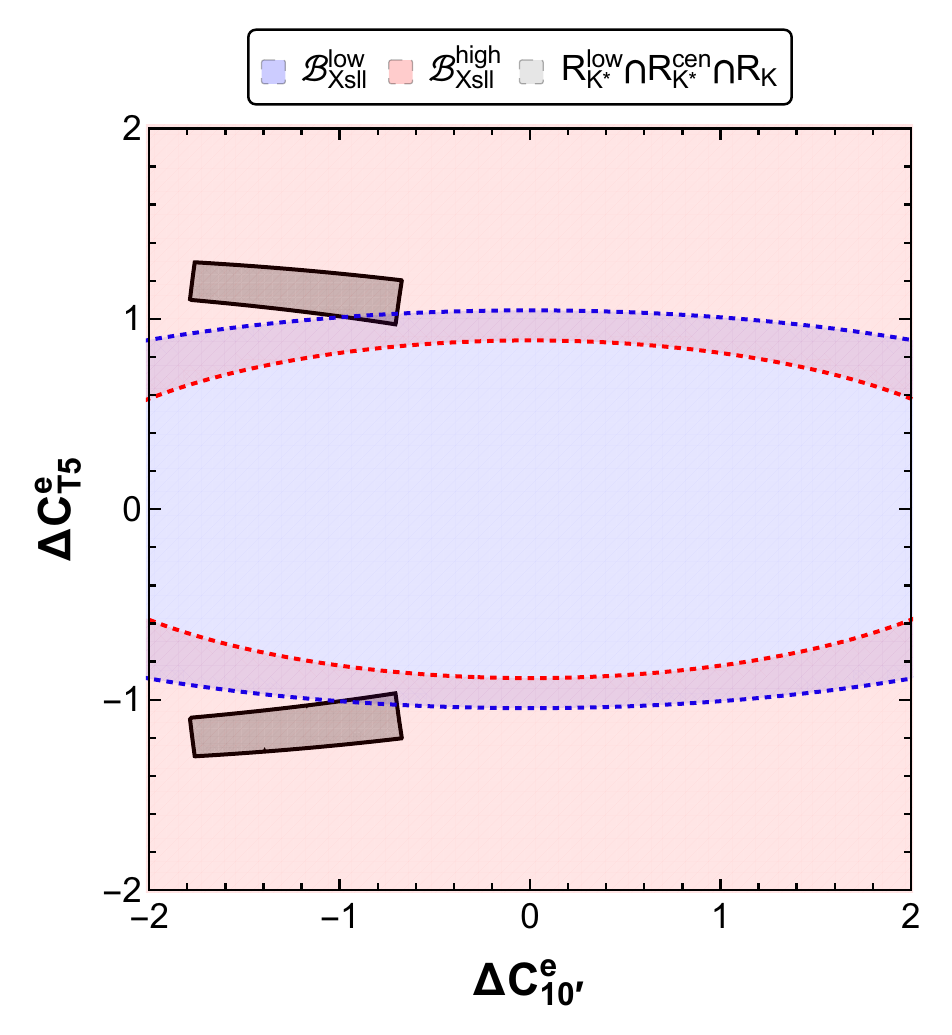}
\end{tabular}
\caption{Allowed regions in $\Delta C_{10'}^e$ - $\Delta C_{T5}^e$ plane. See 
text for more details.
\label{fig:c10pecT5}}
\end{center}
\end{figure}
%%%%%%%%%%%%%%%%%%%%%%%%%%%%%%%%%%%%%%%%%%%%%%%%%%%%%%%%%%%%%%%
Fig.~\ref{fig:c9pecT5}  and Fig.~\ref{fig:c10pecT5} show the allowed regions in 
$\Delta C_{9'}^e$  vs. $\Delta C_{T5}^e$  and 
$\Delta C_{10'}^e$ vs. $\Delta C_{T5}^e$ planes respectively. In these cases 
also, the primed operators can be allowed if a large 
tensor contribution exists at the same time. 

%%%%%%%%%%%%%%%%%%%%%%%%%%%%%%%%%%%%%%%%%%%%%%%%%%%%%%%%%%%%%%%%%%%%%%%%%%%
%\vspace*{-5mm}
\section{\text{\bf SU(2) $\times$ U(1)$_{\rm Y}$ gauge invariance}}
\label{su2u1}
%\vspace*{-3mm}
%%%%%%%%%%%%%%%%%%%%%%%%%%%%%%%%%%%%%%%%%%%%%%%%%%%%%%%%%%%%%%%%%%%%%%%%%%%

As mentioned in the main text, the tensor operators do not get generated at the 
dimension-6 level if SU(2) $\times$ U(1)$_{\rm Y}$ gauge invariance 
is imposed\footnote{Tensor operators have also been considered in the context of 
the charged current anomalies $R_D$ and $R_{D^*}$, see 
for example \cite{Biancofiore:2013ki,Bardhan:2016uhr}. In that case, however, 
tensor operator can be generated already at the dimension 
6 level \cite{Bardhan:2016uhr}.}. However, they can be generated at the 
dimension-8 level. Here we show a few examples, 
 \begin{eqnarray}
1. ~~ &&  \frac{C_{Y_d Y_\ell}}{\Lambda^4} [\overline{s_R}  \sigma^{\mu\nu} Q_3  
\tilde H] [\overline{e_{\ell R}} \sigma_{\mu\nu} L_\ell \tilde H ]   \nn \\
&& \hspace*{0cm} \to \frac12 C_{Y_d Y_\ell} \frac{v^2}{\Lambda^4} 
[\overline{s_R}  \sigma^{\mu\nu} b_L ] [\overline{e_{\ell R}} \sigma_{\mu\nu} 
e_{\ell L} ] \nonumber \\*
&&  \hspace*{0cm} = \frac14 C_{Y_d Y_\ell} \frac{v^2}{\Lambda^4} \left( {\cal 
O}_T- {\cal O}_{T5} \right)
 \end{eqnarray}

\begin{eqnarray}
2.~~ &&\frac{C_{sLeQ}}{\Lambda^4}[\overline{s_R}   L_\ell \tilde H] 
[\overline{e_{\ell R}} Q_3 \tilde H ]\nonumber\\*
&& \hspace*{-0mm} \to \frac{C_{sLeQ}}{\Lambda^4}\bigg(\frac12  [\overline{s_R}  
Q_3  \tilde H] [\overline{e_{\ell R}} L_\ell \tilde H ] \nn \\
&& \hspace*{2cm} + \frac18 [\overline{s_R} \sigma^{\mu\nu} Q_3  \tilde H] 
[\overline{e_{\ell R}} \sigma_{\mu\nu} L_\ell \tilde H ]\bigg)\nonumber\\*
&& = \frac18 C_{sLeQ} \frac{v^2}{\Lambda^4}\left( {\cal O}_{S'} -  {\cal O}_{P'} 
+ \frac14 {\cal O}_{T} - \frac14 {\cal O}_{T5} \right) \label{eq:tensor2}
\end{eqnarray}
%%%%%%%%%%%%%%%%%%%%%%%%%%%%%%%%%%%%%%%%%%%%%%%%%%%%%%%%%%%%%%%
\begin{figure}[htb!]
\begin{center}
\begin{tabular}{cc}
\includegraphics[scale=0.45]{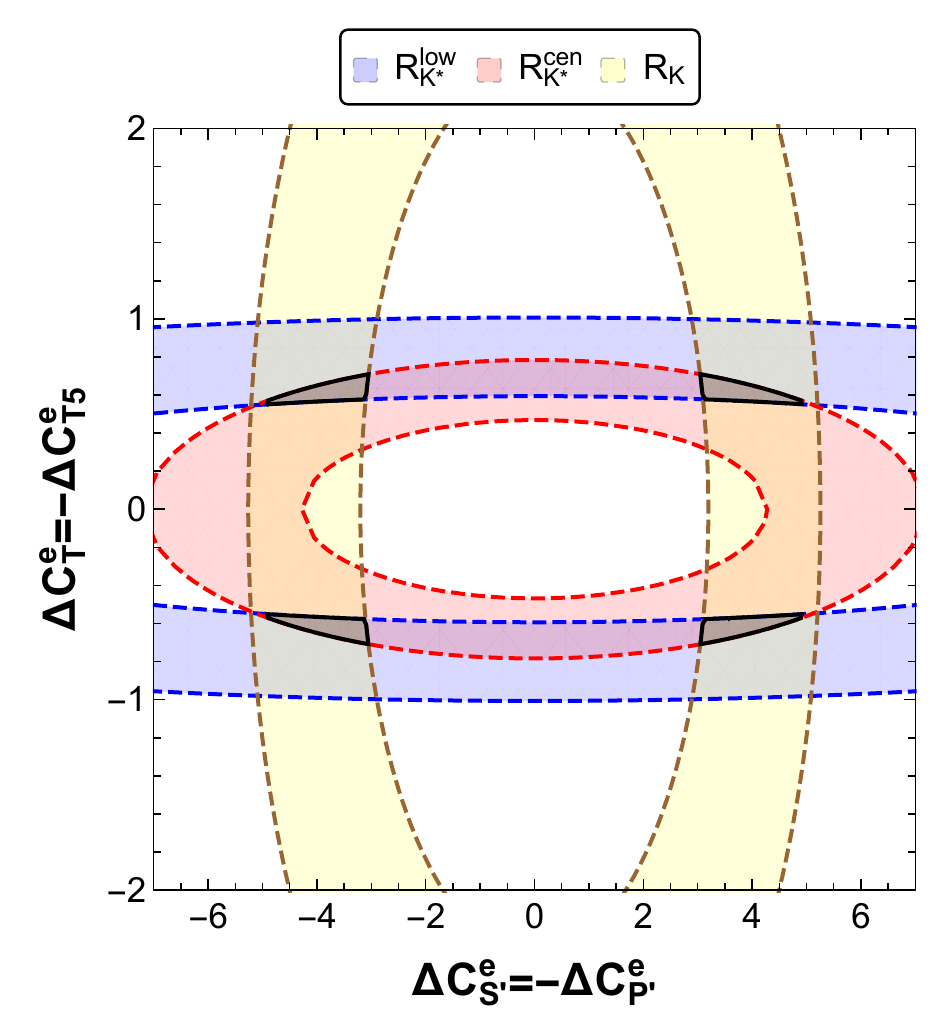} & 
\includegraphics[scale=0.45]{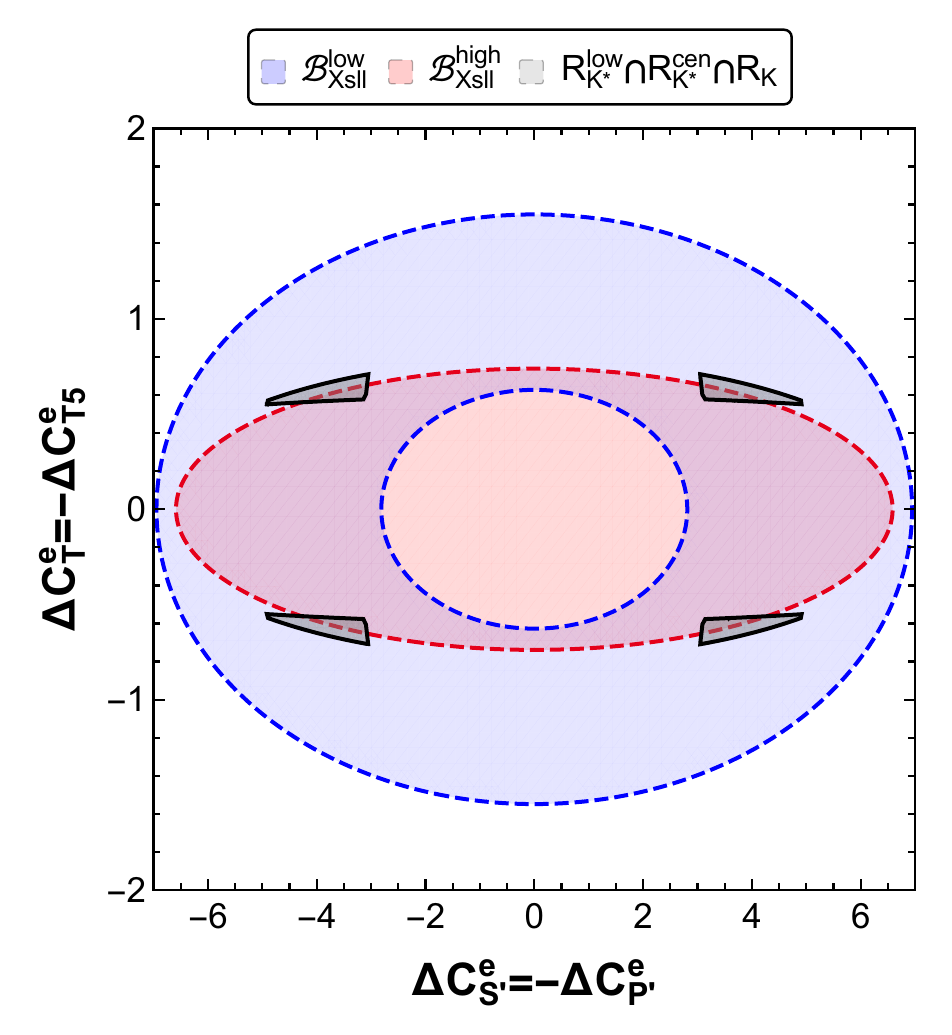}
\end{tabular}
\caption{Allowed regions in $\Delta C_{S'}^e (= - \Delta C_{P'}^e)$ vs.  $\Delta 
C_{T5}^e (= - \Delta C_T^e)$ plane. 
See text for more details.
\label{fig:dim8}}
\end{center}
\end{figure}
%%%%%%%%%%%%%%%%%%%%%%%%%%%%%%%%%%%%%%%%%%%%%%%%%%%%%%%%%%%%%%%
It is hard to generate only the tensor operators in a complete field theory 
model. The second operator above is much easier to generate 
(it can be generated even at the tree level). In this case, however, both scalar 
and tensor operators are generated with the following relations 
among the Wilson coefficients,
\bea
\Delta C_{S'}^e = - \Delta C_{P'}^e = 4 \Delta C_{T5}^e = - 4 \Delta C_T^e \, . 
\label{eq:dim8}
\eea
Note that, gauge invariance at the dimension 6 level 
always leads to the relation $\Delta C_{S'}^e = + \Delta C_{P'}^e$ \cite{Alonso:2014csa}, which is now 
broken by the dimension 8 operators.
In Fig.~\ref{fig:dim8}, we show the various allowed regions in the $\Delta 
C_{S'}^e (= - \Delta C_{P'}^e)$ vs.  $\Delta C_{T5}^e (= - \Delta C_T^e)$ 
plane. It is interesting that the black overlap regions in the left panel 
satisfy Eq.~\eqref{eq:dim8} approximately. In fact, there is tiny region 
in the right panel which satisfies the inclusive measurements too.

Note that, the value of $\Delta C_{S'}^e = - \Delta C_{P'}^e \approx 3$ 
corresponds 
to a NP scale $\Lambda \sim (C_{sLeQ})^{1/4} \, 1.5~\rm TeV$. While the scale is 
rather low, it is still intriguing that one local operator in 
Eq.~\eqref{eq:tensor2} can explain all the anomalies (including $R_{K^*}^{\rm 
low}$) simultaneously. Unfortunately, for such large value 
$\Delta C_{S'}^e = - \Delta C_{P'}^e \approx 3$, $\mathcal{B}_{ee}$ exceeds the 
experimental upper bound, and some cancellation, either 
from other dimension-8 operators or from dimension-6 operators would be 
necessary for this operator to be viable. More detailed 
exploration of such dynamics is left for future work. 

\onecolumngrid
\vspace*{-2mm}
\begin{center}
{\bf 
--------------------------------------------------------------------------------
}
\end{center}

\twocolumngrid
%%%%%%%%%%%%%%%%%%%%%%%%%%%%%%%%%%%%%%%%%%%%%%%%%%%%%%%%%%%%%%%%%%%%%%%%%%%%%%%
%\bibliographystyle{h-physrev5}
%\newpage
%\bibliographystyle{h-elsevier3}
%\bibliography{refs}

%%%%%%%%%%%%%%%%%%%%%%%%%%%%%%%%%%%%%%%%%%%%%%%%%%%%%%%%%%%%%%%%%%%%%%%%%%%%%%%
%%%%%%%%%%%%%%%%%%%%%%%%%%%%%%%%%%%%%%%%%%%%%%%%%%%%%%%%%%%%%%%%%%%%%%%%%%%%%%%

%%%%%%%%%%%%%%%%%%%%%%%%%%%%%%%%%%%%%%%%%%%%%%%%%%%%%%%%%%%%%%%%%%%%%%%%%%%%%%%
\end{document}